\documentclass[lettersize,journal]{IEEEtran}
\usepackage{xcolor,soul,framed} 
\usepackage{cite}
\usepackage{amsmath,amssymb,amsfonts}
\usepackage{graphicx}
\usepackage{textcomp}
\usepackage{dcolumn}
\usepackage{bm}
\usepackage{subfigure} 
\usepackage{pict2e}
\usepackage{epstopdf}
\usepackage{arydshln}
\usepackage{gensymb}
\usepackage{enumitem}
\usepackage[mathscr]{euscript}
\usepackage{stfloats}
\usepackage{makecell}
\usepackage{pifont}

\definecolor{bluerevision}{RGB}{0,112,192}
\newcommand{\blue}[1]{\textcolor{black}{#1}}

\newcommand{\E}{\mathrm{e}}
\newcommand{\jj}{\mathrm{j}}
\newcommand{\cmark}{\ding{51}}%
\newcommand{\xmark}{\ding{55}}%

\makeatletter
 \let\old@ps@headings\ps@headings
 \let\old@ps@IEEEtitlepagestyle\ps@IEEEtitlepagestyle
 \def\confheader#1{%
 \def\ps@headings{%
 \old@ps@headings%
 \def\@oddhead{\strut\hfill#1\hfill\strut}%
 \def\@evenhead{\strut\hfill#1\hfill\strut}%
 }%
 \def\ps@IEEEtitlepagestyle{%
 \old@ps@IEEEtitlepagestyle%
 \def\@oddhead{\strut\hfill#1\hfill\strut}%
 \def\@evenhead{\strut\hfill#1\hfill\strut}%
 }%
 \ps@headings%
 }
 \makeatother
\confheader{%
This work has been accepted for publication in \textit{IEEE Transactions on Wireless Communications}. DOI: 10.1109/TWC.2023.3268949
}
 \usepackage[pscoord]{eso-pic}
\newcommand{\placetextbox}[3]{
 \setbox0=\hbox{#3}
 \AddToShipoutPictureFG*{ \put(\LenToUnit{#1\paperwidth},\LenToUnit{#2\paperheight}){\vtop{{\null}\makebox[0pt][c]{#3}}}
 }
 }
 \placetextbox{.5}{0.055}{\small{\copyright 2023 IEEE. Personal use of this material is permitted. Permission from IEEE must be obtained for all other uses, in any current or future media,}}
 \placetextbox{.5}{0.045}{\small{including reprinting/republishing this material for advertising or promotional purposes, creating new collective works, for resale or redistribution to}}
 \placetextbox{.5}{0.035}{\small{servers or lists, or reuse of any copyrighted component of this work in other works.}}

\begin{document}

\title{Joint Direction-of-Arrival and Time-of-Arrival Estimation with Ultra-wideband Elliptical Arrays}

\author{Alejandro Ramírez-Arroyo, Antonio Alex-Amor, Pablo Padilla, and Juan~F.~Valenzuela-Valdés 
\thanks{This work was supported in part by the Spanish Government under Project PID2020-112545RB-C54, and Project TED2021-129938B-I00; in part by “Junta de Andalucía” under Project A-TIC-608-UGR20, Project P18.RT.4830, and Project PYC20-RE-012-UGR; and in part by the Predoctoral Grant FPU19/01251. (\textit{Corresponding author: Alejandro Ramírez-Arroyo}.)}
\thanks{Alejandro Ramírez-Arroyo, Pablo Padilla and Juan F. Valenzuela-Valdés are with the Department of Signal Theory, Telematics and Communications, Universidad de Granada (UGR), 18071 Granada, Spain (e-mail: alera@ugr.es; pablopadilla@ugr.es;
juanvalenzuela@ugr.es).}
\thanks{Antonio Alex-Amor is with the Department of Information Technologies, Universidad San Pablo-CEU, CEU Universities,  Campus Montepríncipe, 28668 Boadilla del Monte (Madrid), Spain  (e-mail: antonio.alexamor@ceu.es).}

}


\markboth{Joint Direction-of-Arrival and Time-of-Arrival Estimation with Ultra-wideband Elliptical Arrays}%
{Joint Direction-of-Arrival and Time-of-Arrival Estimation with Ultra-wideband Elliptical Arrays}

\IEEEpeerreviewmaketitle

\maketitle


\vspace{-14truemm}

\begin{abstract}
This paper presents a general technique for the joint Direction-of-Arrival (DoA) and Time-of-Arrival (ToA) estimation in multipath environments. The proposed ultra-wideband technique is based on phase-mode expansions and the use of nearly frequency-invariant elliptical arrays. New possibilities open with the present approach, as not only elliptical, but also circular and linear (highly flattened) arrays can be considered with the same implementation. Systematic selection/rejection of signals-of-interest/signals-not-of-interest in smart wireless environments is possible, unlike with previous approaches based on circular arrays. Concentric elliptical arrays of many sizes and eccentricities can be jointly considered, with the subsequent improvement that entails in DoA and ToA detection. This leads to the realization of pseudo-random array patterns; namely, quasi-arbitrary geometries created from the superposition of multiple elliptical arrays. Some simulation and experimental tests (measurements in an anechoic chamber) are carried out for several frequency bands to check the correct performance of the method. The method is proven to give accurate estimations in all tested scenarios, and to be robust against noise and position uncertainty in sensor placement.
\end{abstract}


\newcommand*{\bigs}[1]{\vcenter{\hbox{\scalebox{2}[8.2]{\ensuremath#1}}}}

\newcommand*{\bigstwo}[1]{\vcenter{\hbox{ \scalebox{1}[4.4]{\ensuremath#1}}}}

\section{Introduction}

\IEEEPARstart{D}{irection}-of-arrival (DoA) estimation has been one of the main lines of research in the field of wireless communications in the last decades \cite{DOA1, DOA2, DOA3, DOA4, DOA5, R2_1}. Knowing the position (and time) at which a wave arrives is essential in mobile networks \cite{mobile1, mobile2}, vehicular networks \cite{vehicular1, vehicular2}, MIMO systems \cite{R2_2},  tracking and navigation systems such as GPS \cite{GPS1, tracking1, GPS2}, radar \cite{radar1, radar2}, sonar \cite{sonar}, and many other wireless systems. Even more exotic applications such as failure detection in electronic components are starting to use DoA techniques as a novel alternative \cite{DOA_electronics1, DOA_electronics2}. In fact, new communication environments are materializing these days; namely, vehicle-to-everything (V2X),  ship-to-ship (S2S), high speed train-to-train or UAV-to-UAV \cite{v2x, s2s, t2t, uav, R2_3}. Thus, a great deal of effort is being put into characterizing their most important parameters and \emph{key performance indicators}, direction of arrival (DoA) and time of arrival (ToA) among them, with the aim of improving bandwidth, latency, data rate, power consumption and reliability in present and future communication systems \cite{Alex_IA, Alex_6G, DOA_propagation1, TimeGating, Fan_2016, DOA_propagation2, Fan_2020_2}.

Conventional DoA and ToA estimation methods, i.e., delay-and-sum beamforming \cite{das} or the more classical implementations of MUSIC \cite{DOA1}, ESPRIT \cite{esprit} and maximum-likehood \cite{maximum_likehood} algorithms, were originally developed to work under a narrowband assumption. Naturally, their use is not adequate for today's broadband communication channels such as the high-frequency millimeter-wave links (26, 38, and 60 GHz) dedicated to 5G wireless networks. Thus, many broadband DoA methods have emerged in recent years to overcome the limitations of narrowband approaches \cite{DOA_broadband}. Among the well-established broadband methods, we can find narrowband decomposition (a wideband channel is decomposed into small bins that are treated independently with narrowband techniques) \cite{narrowband_decomposition} or the use of tapped-delay filters with adaptive coefficients \cite{tapfilter}. Although they are simple solutions, both demand high computational resources since the number of required filters increases as the considered bandwidth does. 

Alternatively, \emph{frequency-independent} beamformers (FIB) \cite{fib1, fib2} were formulated to reduce the number of required filters in a fixed bandwidth, and thus the computational demands. Originally, FIB focused on the use of linear arrays. Then, FIB-based algorithms were smartly combined with the omnidirectional characteristics that uniform circular arrays offer to develop efficient methods for DoA and ToA estimation \cite{phase_mode2002}. The outcomes of this original work were promising, as the spatial response of the filter was equalized with a remarkably reduced number of weight coefficients. Nonetheless, some important aspects of the method were later clarified in \cite{phase_mode2008}. Subsequently,  \cite{phase_mode2002} was rapidly extended in different ways. For instance, in \cite{phase_mode2007}, it was demonstrated that the bandwidth of FIB circular arrays can be  broadened and the precision on the DoA estimation improved by employing multiple concentric arrays instead of a single one. In \cite{Zhang_2017, Fan_2019}, multipath components and spherical 3-D propagation were also considered. Moreover, the approach was directly applied to the millimeter-wave frequency range. 

In this document, we propose a generalization of the method successfully developed in \cite{phase_mode2002, phase_mode2007, phase_mode2008, Zhang_2017, Fan_2019} for the joint estimation of the direction-of-arrival and time-of-arrival in multipath environments. Those previous works were based on the use of wideband circular arrays. Here, we demonstrate that \emph{wideband elliptical arrays} can extent the capabilities of the joint estimation in many forms: 
\begin{enumerate}[label=(\roman*)]
    \item Circular geometries are subcases contained by more general elliptical shapes. Thus, we present one of the few methods reported in the literature that is able to deal with circular and elliptical arrays in a single stroke. Moreover, linear arrays can be approximated and analyzed as highly-flattened elliptical arrays. 
    
    \item Elliptical arrays present new degrees of freedom compared to circular arrays. Particularly, the  ellipse radius is not fixed, which provides many configurations for several arrangements based on different eccentricities and rotation angles.

    \item Highly flattened elliptical arrays (and so, linear arrays) show an excellent directivity along the direction of its semi-major axis, as well as remarkable estimation performance for many elevation angles in this direction. Thus, the present methodology allows
    for the efficient and systematic selection of signals-of-interest (SOI) and rejection of signals-not-of-interest (SNOI) in smart radio environments, unlike conventional FIB-DoA methods implemented with circular arrays.
    
    \item Concentric elliptical arrays (CEAs) can be analyzed with the present formulation. As reported by previous works, the use of concentric arrays is expected to significantly improve the accuracy in the joint estimation as well as the bandwidth \cite{phase_mode2007}. Furthermore, the use of concentric elliptical arrays can lead to the implementation of \emph{pseudo-random} grids, a fact that will be exploited in Section~III.D for DoA and ToA estimation. These pseudorandom grids have the advantage of not being limited to a certain number of sensor arrangements.
    
    \item The method is proven to be robust to noise and position uncertainty in DoA and ToA detection. In addition, its computational demands are comparable to state-of-art approaches. 
\end{enumerate}   

The document is organized as follows. Section~II presents the theoretical framework for the joint DoA and ToA estimation with ultra-wideband elliptical arrays. Furthermore, we discuss on the advantages and limitations of the method.  Section~III presents some numerical simulation results in order to validate the present approach. Section~IV presents some experimental results extracted from the measurement facilities at the University of Granada. Finally, conclusions are drawn in Section~V. 

\section{\label{sec:Theory} Theoretical Framework}

\begin{figure}[t]
	\centering
	\includegraphics[width= 0.7\columnwidth]{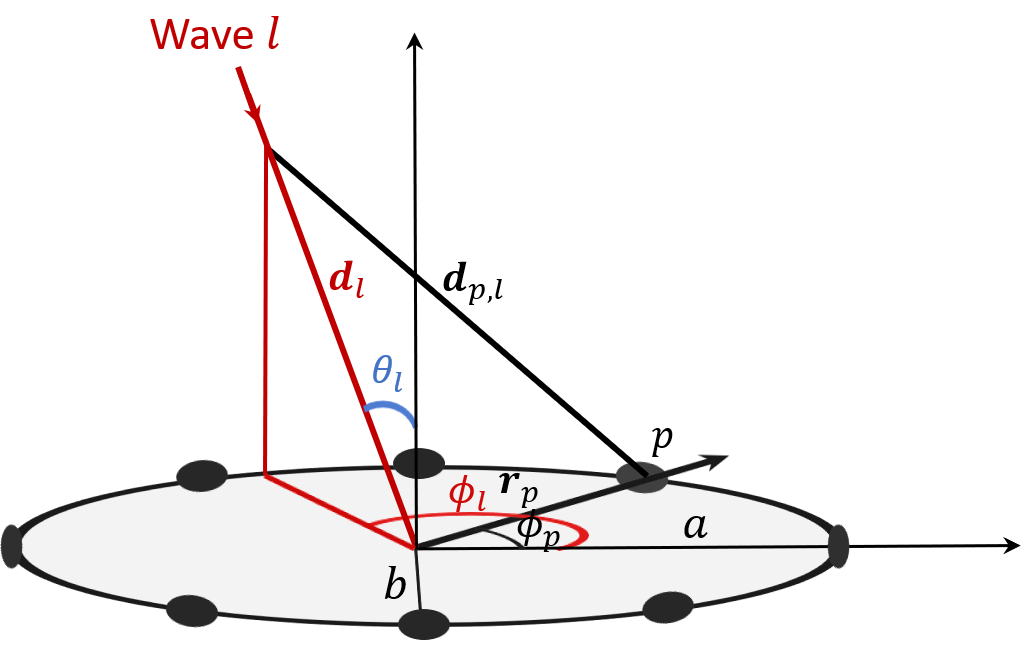}
	\caption{Scheme illustrating the different parameters involved in the joint estimation of DoA and ToA with a single elliptical array.} 
	\label{fig1}
\end{figure}

Let us consider the situation illustrated in Fig. \ref{fig1}. An incident spherical wave $l$, described by the \emph{unknown} azimuth and elevation  angles $\phi_l$ and $\theta_l$, respectively, impinges on the $P$ sensors ($p=0, 1, ... ,P-1 $) that conform the elliptical array. The elliptical array is defined by its eccentricity $\xi = \sqrt{1 - b^2/a^2}$, which relates the semi-major ($a$) and semi-minor ($b$) axes, respectively. The frequency response at the center of the elliptical array is given by 
\begin{equation}
    H_l(f) = \kappa_l\, \E^{\jj 2 \pi f \tau_l}\, ,
\end{equation}
where $\kappa_l$ is the attenuation for path/wave $l$ for any given path loss exponent, and $\tau_l$ stands for the propagation delay for path/wave $l$. Both parameters are to be determined. 

The frequency response at the $p$-th sensor, $H_{p, l}(f)$, will include a phase shift with respect to the frequency response estimated at the center of the array,  $H_l(f)$. By applying trigonometric relations, we arrive to
\begin{equation} \label{Hpl}
H_{p, l}(f)=\left(\sqrt{\frac{d_{\ell}}{d_{p, \ell}}}\right)^{\;\gamma} H_{\ell}(f)\, \E^{\jj 2 \pi f \Delta d_{p, \ell} / c}\, ,
\end{equation}
where $c$ is the speed of light, the term $\left(\sqrt{d_{\ell}/d_{p, \ell}}\right)^{\;\gamma}$ accounts for amplitude attenuation given a path loss exponent $\gamma$ between the center of the array and the $p$-th sensor, and
\begin{equation}
    \Delta d_{p, l} = d_l - d_{p,l}\, ,
\end{equation}
with
\begin{equation}
    d_{p,l} = ||\mathbf{d}_l - \mathbf{r}_p|| = \sqrt{d_l^2 + r_p^2 - 2d_l r_p \sin (\theta_l) \cos(\phi_l - \phi_p)}\, .
\end{equation}
%

\noindent In the former expression, $d_l$ is the distance from the center of the elliptical array to the source point, $r_p$ is the distance from the center of the  array to sensor $p$, $\phi_p$ is the azimuth angle associated to sensor $p$ (see Fig. \ref{fig1}), both given by
\begin{equation}  \label{rp}
    r_p = \sqrt{x_p^2 + y_p^2}\, ,
\end{equation}
\begin{equation}
    \phi_p = \arctan{\left( \frac{y_p}{x_p} \right)}\, .
\end{equation}
Additionally, the Taylor series for $\Delta d_{p,l}$ allows to expand this term as \cite{Fan_2019}
\begin{equation}
    \Delta d_{p,l} \approx r_p \sin (\theta_l) \cos(\phi_l - \phi_p)\, .
\end{equation}
Thus, \eqref{Hpl} can be expressed as
\begin{equation} \label{Hpl2}
    H_{p, l}(f)=\left(\sqrt{\frac{d_{\ell}}{d_{p, \ell}}}\right)^{\;\gamma} H_l(f)\, \E^{\jj 2 \pi f r_p \sin (\theta_l) \cos(\phi_l - \phi_p) / c}\, .
\end{equation}
Considering that the source $l$ is located far from the elliptical array, the term $d_l/d_{p, l} \approx 1$, so \eqref{Hpl2} can be simplified to
\begin{equation} \label{Hpl3}
    H_{p, l}(f)= H_l(f)\, \E^{\jj 2 \pi f r_p \sin (\theta_l) \cos(\phi_l - \phi_p) / c}\, .
\end{equation}
Now, we particularize the study to the case $\theta_l = 90^\mathrm{o}$; namely, where the incident plane coincides with the plane where the elliptical array is located. Thus, \eqref{Hpl3} reduces to
\begin{equation} \label{Hpl4}
    H_{p, l}(f)= H_l(f)\, \E^{\jj 2 \pi f r_p \cos(\phi_l - \phi_p) / c}\, .
\end{equation}
This approximation is fundamental in order to develop further steps of the method. Although it may seem prohibitive, this approach was demonstrated to work  over a wide range of elevation angles different from $\theta_l = 90^\mathrm{o}$ \cite{Fan_2019}.

The Jacobi-Anger identity \cite{Jacobi-Anger} allows us to expand the former expression and \emph{decouple} the frequency-dependent and phase-dependent components in two different terms as
\begin{equation} \label{Hpl5}
    H_{p, l}(f)= H_l(f) \sum_{n=-\infty}^\infty \jj^n J_{n,p}\left(\frac{2\pi f r_p}{c}\right) \, \E^{\jj n (\phi_l - \phi_p) }\, ,
\end{equation}
where $J_{n,p}(\cdot)$ is the Bessel function of the first kind of order $n$ associated to sensor $p$. At a first glance, some differences can be observed with respect to the cases shown in \cite{phase_mode2007, phase_mode2008, Fan_2019} for circular arrays. When considering elliptical arrays, the frequency-dependent component [$J_{n,p}(\cdot)$] varies for each sensor $p$. This is due to the fact that the distance between the center of the array and the position of sensor $p$ is not constant [$r_p = r_p(\phi_p) \neq r$], unlike in circular arrays where this distance is the same for all sensors ($r_p = r$).

The azimuth angle $\phi_l$ can be estimated by applying a phase-mode expansion, $H_{m,l}(f)$,  \cite{phase_mode2007, phase_mode2008, Fan_2019} to the former expression that includes basis functions of the form $\E^{\jj m \phi_p}$ ($m$~being the integer order of the phase-mode) and a frequency-dependent filter $W_{m,p}(f)$, leading to
\begin{multline} \label{Hml1}
\begin{aligned}
   H_{m,l}(f) &=  \frac{1}{P} \sum_{p=0}^{P-1} H_{p, \ell}(f)\, \E^{\jj m \phi_{p}}\, W_{m,p}(f)\\ 
   & = \frac{1}{P} H_{\ell}(f)
   \sum_{p=0}^{P-1} \,
   \sum_{n=-\infty}^{+\infty} \jj^n
    J_{n,p}\left(2 \pi f \frac{r_{p}}{c}\right) W_{m,p}(f)\\
    & \hspace{3cm} \times\; e^{\jj n \phi_{\ell}}  \, e^{\jj(m-n) \phi_{p}}\, .
\end{aligned}
\end{multline}
The infinite sum in $n$ can be split in two different addends: the term $n = m$ and a series for all $n \neq m$. This allows to rewrite \eqref{Hml1} as
\begin{multline} \label{Hml2}
   H_{m,l}(f) = \frac{1}{P} H_{\ell}(f)
   \sum_{p=0}^{P-1} W_{m,p}(f) \\ 
     \hspace{-3cm} \times \Big[ \jj^m J_{m,p}\left(2 \pi f \frac{r_{p}}{c}\right)  \E^{\jj m \phi_{\ell}} \\
     + \sum_{\substack{n=-\infty \\ n\neq m}}^{+\infty}  \jj^n 
J_{n,p}\left(2 \pi f \frac{r_{p}}{c}\right)
 \E^{\jj n \phi_{\ell}}  \, \E^{\jj(m-n) \phi_{p}} \Big].
\end{multline}
Previous works \cite{phase_mode2007, phase_mode2008, Fan_2019} focused on circular arrays took advantage of the independence of the Bessel functions with respect to $p$ [$J_{n,p}(\cdot) = J_{n}(\cdot)$] and the orthogonality relations of the sum $\sum_{p=0}^{P-1} \E^{\jj(m-n) \phi_{p}}/P$ to cancel out the infinite series ($\forall n \neq m$) in eq. \eqref{Hml2}. However, this is not exactly the situation for elliptical arrays. Thus, some approximations have to be considered in order to proceed. Under the assumption of low eccentricity $\xi$ levels (ellipses that are not too flattened, $\xi \ll 1$), the former discussion still applies since $r_p \approx r$ (constant radius in average) and the sensors in the elliptical array are quasi-uniformly distributed ($\phi_p \approx 2\pi p/P$). As a consequence, $J_{n,p}(2\pi f r_p/c) \approx J_{n}(2\pi f r/c)$ can be cleared out from the sum in $p$ and the orthogonality relation
$\sum_{p=0}^{P-1} \E^{\jj(m-n) \phi_{p}}/P = 0\,\,, \forall n\neq m$, remains valid. Thus, the last double sum in eq.~\eqref{Hml2} is approximately zero under low-eccentricity assumption; namely, 
\begin{multline}
   \frac{1}{P}\sum_{p=0}^{P-1} W_{m,p}(f) \sum_{\substack{n=-\infty \\ n\neq m}}^{+\infty}  \jj^n 
J_{n,p}\left(2 \pi f \frac{r_{p}}{c}\right)
 \E^{\jj n \phi_{\ell}}  \, \E^{\jj(m-n) \phi_{p}} \\ \stackrel{\xi \ll1}{\approx} 
 W_{m}(f) \sum_{\substack{n=-\infty \\ n\neq m}}^{+\infty} \jj^n J_{n}\left(2 \pi f \frac{r}{c}\right)  \E^{\jj n \phi_{\ell}} \sum_{p=0}^{P-1} \frac{\E^{\jj(m-n) \phi_{p}}}{P}  = 0\, .
\end{multline}
Thus, eq. \eqref{Hml2} is simplified to
\begin{equation} \label{Hml3}
   H_{m,l}(f) \approx \frac{\jj^m}{P}   H_{\ell}(f) \,\E^{\jj m \phi_{\ell}}\,   \, \sum_{p=0}^{P-1} J_{m,p}\left(2 \pi f \frac{r_p}{c}\right) W_{m,p}(f).
\end{equation}
The approximation taken above, although originally derived for low eccentricities, will be demonstrated in Sections III and IV to be applicable to a wide range of elliptical arrays, even those that reach eccentricity values of $\xi = 0.99$, as long as the number of considered sensors is large. The reason is that, even for high eccentricities, the main contribution of the transformation is given by the phase mode $n = m$, while the rest of the contributions (sum $\forall n \neq m$) is normally negligible [see eq.(13)]. Nonetheless, an increment of the artifact levels is expected as the eccentricity increases, thus degrading the joint DoA and ToA estimation in some angular regions. In practice, $P$ should be selected so the sensor separation is less than $\lambda/2$, where $\lambda$ stands for the wavelength.

Then, we can reduce \eqref{Hml2} to
\begin{equation} \label{Hml4}
    H_{m,l}(f) \approx H_{\ell}(f) \,\E^{\jj m \phi_{\ell}} \, ,
\end{equation}
if we consider the filter $W_{m,p}(f)$ to be 
\begin{equation} \label{Wm1}
    W_{m,p}(f) = \frac{1}{\jj^m\, J_{m,p}\left(2 \pi f \frac{r_p}{c}\right)} \, .
\end{equation}
Alternatively, we can choose $W_{m,p}(f)$ as
\begin{equation} \label{Wm2}
W_{m,p}(f)= \frac{2} {\jj^{m}\left[J_{m,p}(2 \pi f r_{p} / c)- \jj\, J_{m,p}^{\prime}(2 \pi f r_{p} / c)\right]}\, ,
\end{equation}
since it was proven in \cite{Fan_2019} to give more accurate results in realistic scenarios where elevation angles different from $\theta_l =\nolinebreak90^\mathrm{o}$ were involved. Additionally, eq.~\eqref{Wm1} involves deep nulls for some frequencies $f$, a fact that limits the  channel bandwidth. The situation is different for the filter depicted in eq.~\eqref{Wm2}, whose performance is similar to that in eq.~\eqref{Wm1} but operates over a noticeably larger bandwidth. A detailed discussion on the filter choice can be found in~\cite{Zhang_2017}. In eq.~\eqref{Wm2}, $J_{m,p}^{\prime}(\cdot)$ is the first derivative of the Bessel function of order~$n$ related to sensor~$p$.

The array response for the $m$-th phase mode is calculated by taking into account the contribution of the $L$ incident waves as
\begin{equation} \label{Ym}
    H_m(f) = \sum_{l=1}^L H_{m,\ell}(f) \approx \sum_{l=1}^L H_{\ell}(f) \,e^{\jj m \phi_{\ell}}\, ,
\end{equation}
which admits a matrix-form representation $\mathbf{H}(m,f)$. Finally, the joint DoA and ToA estimation is extracted by computing the 2-D fast Fourier transform (FFT) of $\mathbf{H}(m,f)$,
\begin{equation} \label{Ytilde}
    \widetilde{\mathbf{H}}(\phi, \tau) = \mathrm{FFT}\{\mathbf{H}(m,f)\},
\end{equation}
which includes a general representation of the wireless channel properties for all considered angular and time steps. Note that DoA and ToA resolution of $\widetilde{\mathbf{H}}(\phi, \tau)$ is given by the number of phase modes and frequency samples in $\mathbf{H}(m,f)$. Concretely, DoA resolution is calculated as $360\degree/M$, where $M$ is the number of considered phase modes. Thus, higher DoA resolution requires of a larger number of phase modes. ToA resolution is calculated as $1/B$, where $B$ is the bandwidth of the channel. The maximum observable time in the estimation is $(K-\nolinebreak1)/B$, which is determined by $B$ and the number of frequency samples $K$. Therefore, channel bandwidths belonging to the range of ultra-wideband technologies are required to achieve good spatial resolution in the estimation since there exists a trade-off between the spatial resolution and~$B$.

The maximum number of considered phase modes $M$ [eq.~\eqref{Hml3}] is restricted by the denominator in $W_{m,p}(f)$ [eq.~\eqref{Wm2}]. This denominator may introduce numerical errors as it approaches zero. In order to avoid numerical instabilities, $M$ must be chosen such that
\begin{equation} \label{notapprox}
    \jj^{m}\left[J_{m,p}(2 \pi f r_{p} / c)- \jj\, J_{m,p}^{\prime}(2 \pi f r_{p} / c)\right] \not\approx 0 \quad \forall m.
\end{equation}
To fulfill the previous condition, we know that there is a number of phase modes, namely $\lvert M_{lim} \rvert$, above which the Bessel function $J_{m,p}(\cdot)$ and its derivative tend to zero \cite{phase_mode2008}. Therefore, stability can be guaranteed if $M$ is chosen below $\lvert M_{lim} \rvert$ \cite{Zhang_2017}. This limit value is dependent on the argument of $J_{m,p}(\cdot)$ and $J_{m,p}^{\prime}(\cdot)$. The larger the argument ($2 \pi f r_p / c$), the larger $\lvert M_{lim} \rvert$. Consequently, i) high frequencies provide higher resolution in DoA estimation, and ii) $\lvert M_{lim} \rvert$ is defined by the sensor $p$ whose distance $r_p$ is the smallest for a given frequency $f$. This distance turns out to be the semi-minor axis in elliptical arrays. Throughout Sections III and IV, it is shown that eccentricity values up to 0.99 allow the method to work properly even when $\lvert M_{lim} \rvert$ is reduced.


\subsection{Improving the Efficiency of the Method}

It can be noted from eqs. \eqref{Wm1}, \eqref{Wm2} that each sensor $p$ that conforms the elliptical array has a different filter $W_{m,p}(f)$. Considering $M$ phase modes in the computation, elliptical arrays will require of $M \times P$ filters for the  DoA and ToA estimation. This situation is different in circular arrays, where all the sensors share the same filter for the $m$-th phase mode. Thus, only $M \times 1$ filters are needed in the case of considering circular arrays. This causes that the joint estimation with ultra-wideband elliptical arrays is less computationally efficient than with ultra-wideband circular arrays, despite the multiple benefits that elliptical arrays offer (generality, selectivity, pseudo-random grids, etc.) compared to circular ones. Nonetheless, we can exploit the symmetries of elliptical geometries and Bessel functions to improve the efficiency of the proposed method. 

Let us name as $\alpha_p$ an azimuth angle that is contained in the first quadrant of the elliptical array. Angles of the form $180^\mathbf{o} - \alpha_p$ (second quadrant) , $180^\mathbf{o} + \alpha_p$ (third quadrant), and $360^\mathbf{o} - \alpha_p$ (fourth quadrant) have associated the same value of $r_p$. Therefore, the number of required filters in the joint estimation can be reduced to $M \times [P/4 + 1]$ by exploiting the symmetries of the ellipse. Thus, when $P$ is large, the required number of filters is approximately reduced by a factor of 4 compared to the raw processing. 

We can further reduce the complexity of the problem by considering the symmetry of Bessel functions. Concretely, we can take advantage of the following expression
\begin{equation}
    J_{-m,p}(\chi_p) = (-1)^m J_{m,p}(\chi_p)\, ,
\end{equation}
that relates negative and positive integer orders $m$. Thus, the number of filters would reduce to $[M/2 + 1] \times [P/4 + 1]$. For large values of $M$ and $P$, the number of required filters is asymptotically reduced from $M \times P$ to $M/2 \times P/4$, that is, by a factor of 8.

Additionally, the use of a \emph{single average filter} $\overline{W}_{m}(f)$ can provide good DoA and ToA estimations when low- and medium-eccentricity elliptical arrays are involved. The average filter is given by replacing all $r_p$ values in eqs. \eqref{Wm1}-\eqref{Wm2} by  $r$,  computed as the average between the semi-major and semi-minor axes of the ellipse.  Note that in the case of considering a single average filter, the number of required filters is reduced to $M \times 1$, as in the case of circular arrays. Therefore, a good approximation for elliptical arrays with low and medium eccentricities can be obtained, notably reducing the computational complexity of the problem. The single-average-filter approach is expected to give accurate results as long as the elliptical array is not highly flattened. In practice, this approach has been found to be valid up to $\xi \lesssim0.7$. Finally, note that the time required in the total propagation channel characterization process is limited by the measurement acquisition. The previous fact is true for setups based on virtual arrays. Due to the high number of sensors required to fulfill the spatial Nyquist theorem, phase-mode expansion for DoA and ToA characterization makes use of these virtual arrays \cite{phase_mode2008, Zhang_2017, Fan_2019}. Therefore, the measurement process takes orders of magnitude longer than the time required to apply the estimation method. Thus, the global time for taking measurements and applying the method ends up being similar in both circular and elliptical arrays.

\subsection{Rotated Elliptical Arrays}

\begin{figure}[t]
	\centering
	\subfigure[]{\includegraphics[width=0.4\columnwidth]{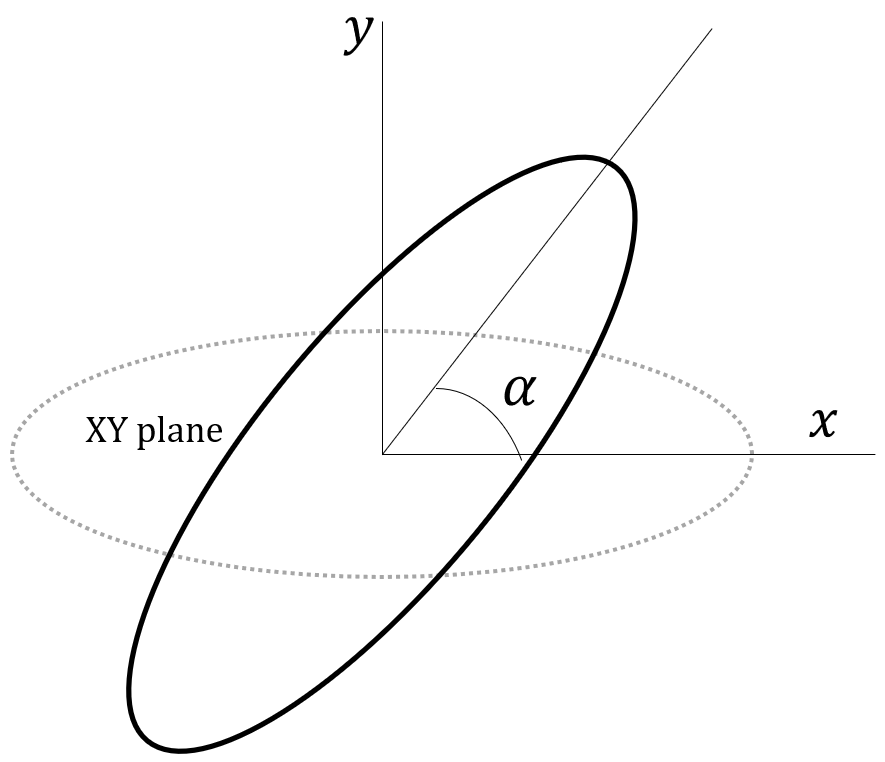}
	} \hspace{0.5cm}
	\subfigure[]{\includegraphics[width= 0.4\columnwidth]{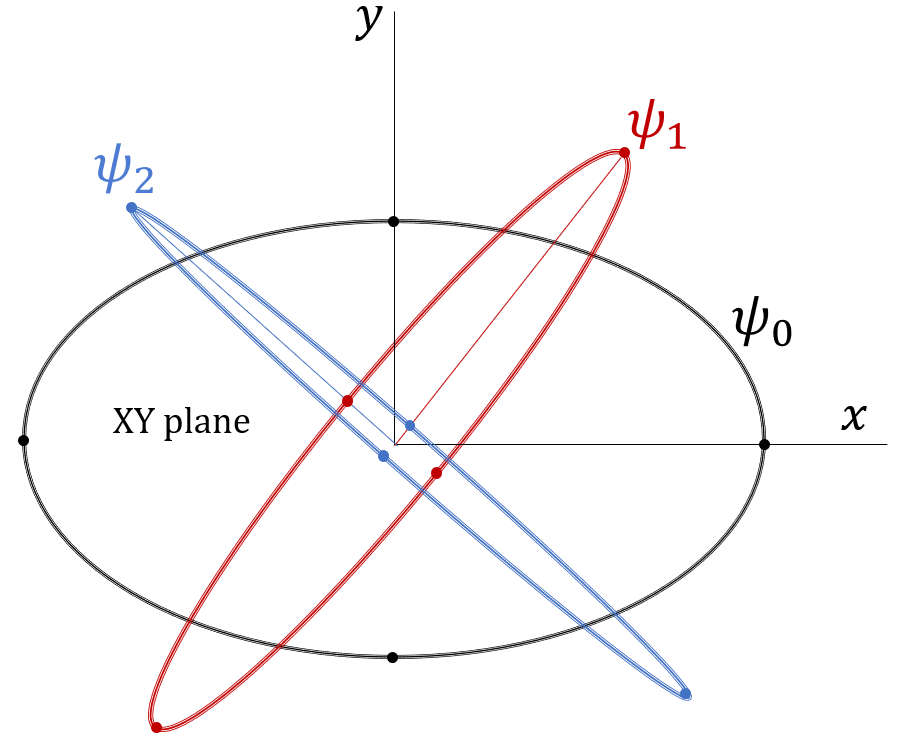}
	}
	\caption{Representation of a (a) single rotated elliptical array and (b) three concentric elliptical arrays. All the arrays are lying in the XY plane.}
	\label{fig2}
\end{figure}

A simple modification can be applied to the present formulation for the convenient use of rotated elliptical arrays in DoA and ToA estimation. Actually, this will serve as the basis for the analysis of advanced scenarios (concentric arrays, pseudo-random grids) in next sections. Considering that the elliptical array is azimuthally rotated counterclockwise by an angle $\alpha$ [see Fig.~\ref{fig2}(a)], the values of $x_p$ and $y_p$ in eq. \eqref{rp} should be simply replaced by
\begin{align}
    \begin{split}
        x_p \rightarrow x_p\cos \alpha + y_p \sin \alpha \\
        y_p \rightarrow x_p\sin \alpha - y_p \cos \alpha 
    \end{split} \quad.
\end{align}

Alternatively, it could be convenient to work with the elliptic coordinate system, formed by the radial coordinate $\rho \in [0, \infty)$ and the angular coordinate $\eta \in [0, 2\pi)$, when positioning the sensors that conform the rotated elliptical array. In that case, the $p$-th sensor will be located at
\begin{align}   \label{eq_rotada}
    \begin{split}
        x_p = a \cos \eta_p \cos \alpha - b \sin \eta_p \sin \alpha \\
        y_p = a \cos \eta_p \sin \alpha + b \sin \eta_p \cos \alpha 
    \end{split} \quad,
\end{align}
where $a$ and $b$ are the semi-major and semi-minor axes of the elliptical array, respectively. Both parameters are related through the eccentricity value $\xi$. Note that a uniform angular placement for the sensors in elliptical coordinates, $\eta$, does not necessarily imply a uniform angular placement in polar coordinates, $\phi_p$.

\subsection{Concentric Elliptical Arrays}

As an additional step, the joint estimation of the direction-of-arrival and time of arrival can be generalized to the case where multiple concentric elliptical arrays are considered [see Fig. \ref{fig2}(b)]. As pointed out in previous works \cite{phase_mode2007},  the fact of including several concentric arrays is expected to increase the accuracy in the estimation, as well as the frequency response of the entire array if the geometry of the concentric elliptical array is appropriately selected. This is because the response of the entire array is a combination of the individual arrays that compose it \cite{phase_mode2007}. Additionally, the use of ultra-wideband concentric elliptical arrays will lead to advanced functionalities in DoA and ToA detection. These advanced functionalities will be deeply explored in Sections III and IV of the present manuscript. 

Let us consider $\Psi$ concentric elliptical arrays ($\psi = 0, 1, ..., \Psi-1$) lying in the same plane (XY plane), each of them constituted by $P$ sensors. The phase-mode expansion for the $\psi$-th elliptical array will be now
\begin{multline} \label{Hml_psi}
\begin{aligned}
   H_{m,l, \psi}(f) &=  \frac{1}{P} \sum_{p=0}^{P-1} H_{p, l, \psi}(f)\, \E^{\jj m \phi_{p, \psi}}\, W_{m,p, \psi}(f) \\
   & = \frac{1}{P} H_{\ell}(f)
   \sum_{p=0}^{P-1} \,
   \sum_{n=-\infty}^{+\infty} \jj^n
J_{m,p, \psi}\left(2 \pi f \frac{r_{p,\psi}}{c}\right) \\ 
 & \hspace{1.5cm} \times\; W_{m,p, \psi}(f)\, e^{\jj n \phi_{\ell}}  \, e^{\jj(m-n) \phi_{p, \psi}}\, .
\end{aligned}
\end{multline}
Thus, the  phase-mode expansion of the whole system, $H_{m, l}(f)$, will be of the form
\begin{equation} \label{Hml_psi2}
\begin{aligned}
H_{m, l}(f) &=\frac{1}{\Psi} \sum_{\psi=0}^{\Psi - 1} H_{m, l, \psi}(f) \approx H_{\ell}(f) \,e^{\jj m \phi_{\ell}}\,,
\end{aligned}
\end{equation}
which is simply the average of the individual phase-mode contributions of each elliptical array. In fact, notice that the resulting expression for $H_{m, l}(f)$ in eq. \eqref{Hml_psi2} is identical to eq. \eqref{Hml4}, despite the computation of the filters $W_{m,p, \psi}(f)$ is different in this case for each elliptical array. 

Naturally, the computational complexity increases when considering concentric arrays. Now, the system is formed by $P \times \Psi$ sensors, so the number of required filters for DoA and ToA estimation increases to $M \times P \times \Psi$. Nonetheless, by following the recommendations given in Sec. II.A, the number of filters can be similarly reduced by a factor of 8.

\section{\label{Results_theory} Simulations}

This Section shows and discusses some results obtained through the formulation derived in Sec.~II for DoA and ToA estimation in ultra-wideband elliptical arrays. Regarding the organization, Sec.~III.A analyzes the accuracy of the method for single elliptical arrays and the influence of eccentricity $\xi$ and rotation angle $\alpha$ in the estimation. Sec.~III.B introduces concentric elliptical arrays and their benefits compared to single elliptical arrays. Sec.~III.C presents the performance of the joint estimation for elevation angles  different than~$90\degree$. Finally, Sec.~III.D explores the use of pseudorandom grids based on concentric elliptical arrays.

\begin{figure}[t]
	\centering
	\includegraphics[width= 1\columnwidth]{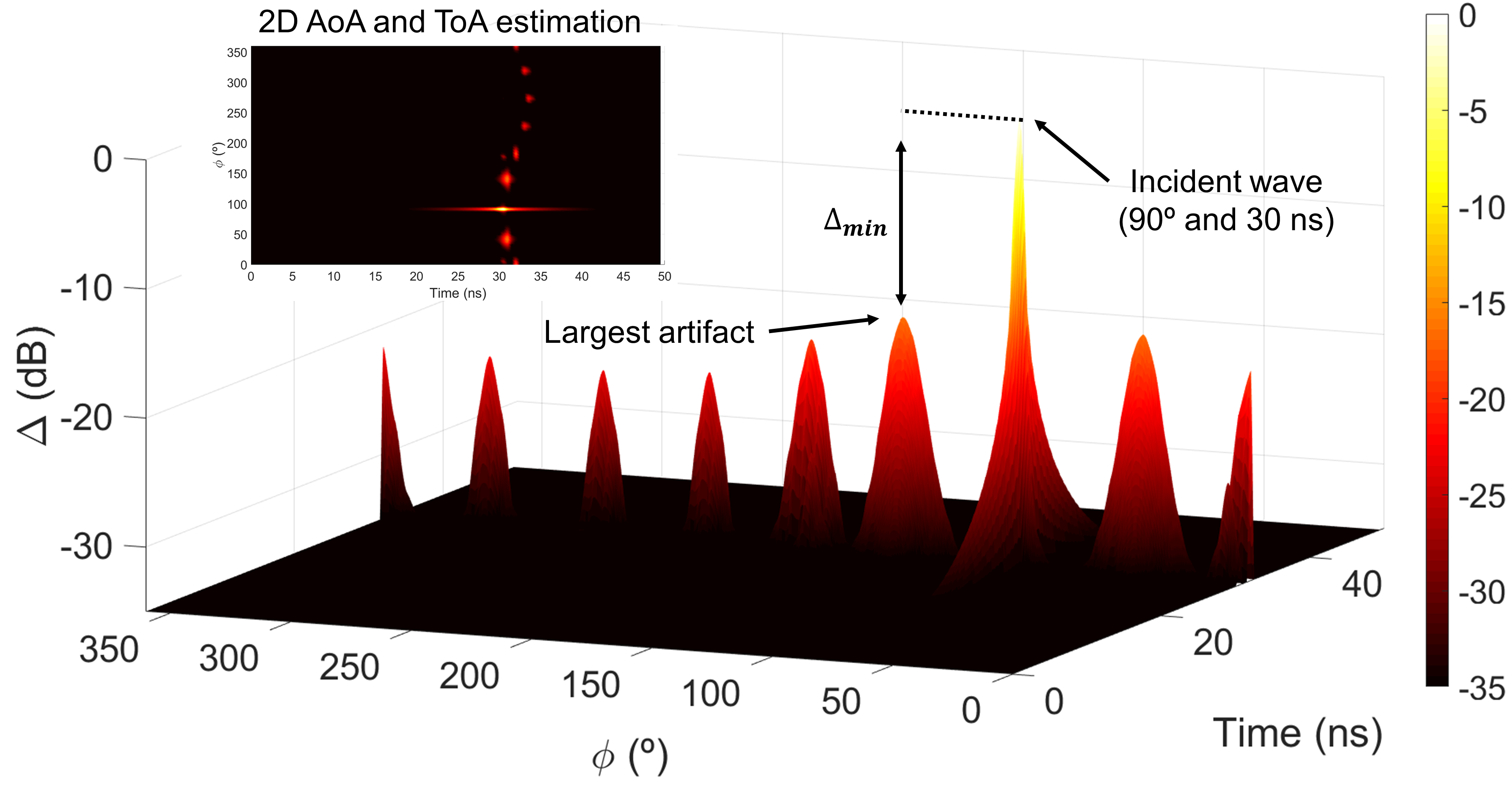}
	\caption{DoA and ToA estimation example for a single incident wave at $\phi_l = 90\degree$ and $\tau_l = 30$ ns. A 2D representation of the angular-temporal domain is shown in the upper left corner.} 
	\label{fig_ejempo_delta}
\end{figure}

In order to validate the estimation for elliptical arrays, it must be ensured that DoA and ToA estimation matches $\phi_l$ and $\tau_l$ for the incident wave, as well as the artifacts (non-desired values expected to be several orders of magnitude below the real path of the incident wave) do not disguise the correct angles and delays. As an example, Fig.~\ref{fig_ejempo_delta} shows the joint angular and delay domain of a particular $\widetilde{\mathbf{H}}(\phi, \tau)$, with a single incident wave at $\phi_{l} = 90\degree$ and $\tau_{l} = 30$~ns. Both parameters are correctly estimated. However, some artifacts can be observed. If they were large compared to the real incident wave, they could misled the real path. Thus, we define~$\Delta$ as the ratio between the correct estimation and the largest artifact in the azimuth and delay domain. The larger the $\Delta$, the better the estimation. Throughout this document, this ratio will be considered as a metric. 

\subsection{Single Elliptical Arrays}

The main benefits of elliptical arrays compared to circular arrays are the new degrees of freedom related to the sensor position. While in circular arrays we only have the control of the radius $r$, three different independent parameters can be modified in elliptical arrays: i) semi-major axis $a$, ii)~eccentricity $\xi$, and iii) rotation angle $\alpha$. This fact allows for a wide variety of sensor arrangements.

The first simulation aims to determine the influence of the eccentricity on the joint DoA and ToA estimation. For this purpose, four ellipses with $\xi = 0$, $0.7$, $0.95$ and $0.99$ are simulated for an incident azimuth range $\phi_l = [-90\degree, 90\degree]$ and $\tau_l = 30$~ns. The frequency band is chosen to be from 28 GHz to 30 GHz ($B = 2$ GHz) for $K = 100$ frequency samples. These frequencies are part of the band n257 defined by the 3rd Generation Partnership Project (3GPP) and are expected to be fundamental in the deployment of 5G New Radio (5G~NR)  \cite{3GPP_TS}.
In order to ensure a proper DoA and ToA estimation, spatial Nyquist theorem must be fulfilled. This implies that the separation between adjacent sensors must be less than half wavelength. In the present work, we have typically assumed a distribution of $P=720$ sensors per array (angular spacing of $0.5\degree$), ensuring that spatial Nyquist theorem is fulfilled for all sensors. Naturally, the number of sensors could be further reduced from 720, normally at the expense of degrading the performance in the estimation. The considered semi-major axis $a = 0.5$ m and the ellipse is placed with its semi-major axis oriented along the horizontal direction ($\alpha = 0\degree$). Given the highest frequency $f = 30$ GHz, we can ensure that the largest separation between sensors is $0.437\lambda$ for any value of eccentricity. Finally, the elevation incident angle  is $\theta_l = 90\degree$, i.e., matching the plane of the elliptical array sensors.

\begin{figure}[!t]
	\centering
	\subfigure[]{\includegraphics[width= 1\columnwidth]{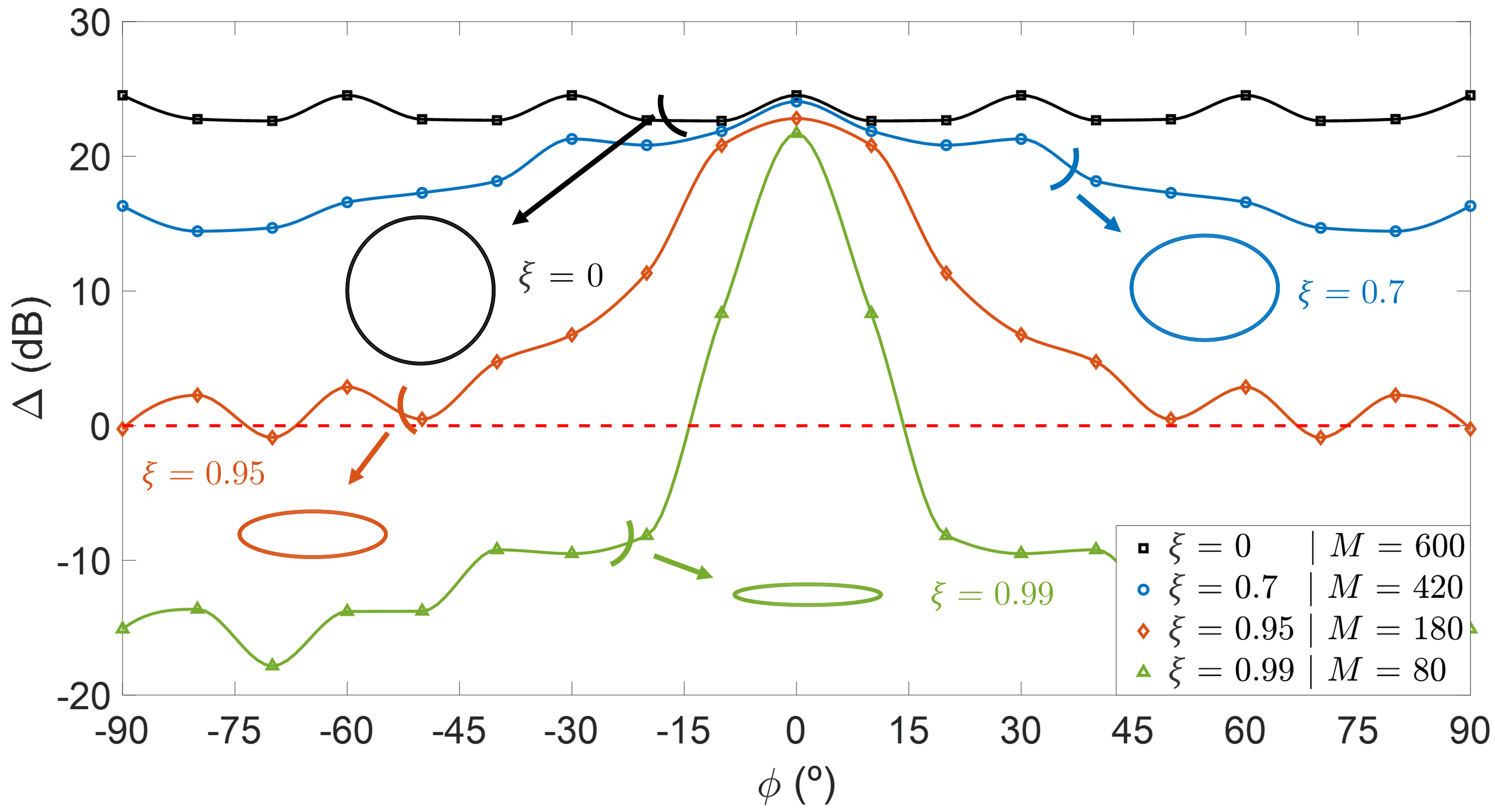}
	}
	\subfigure[]{\includegraphics[width= 1\columnwidth]{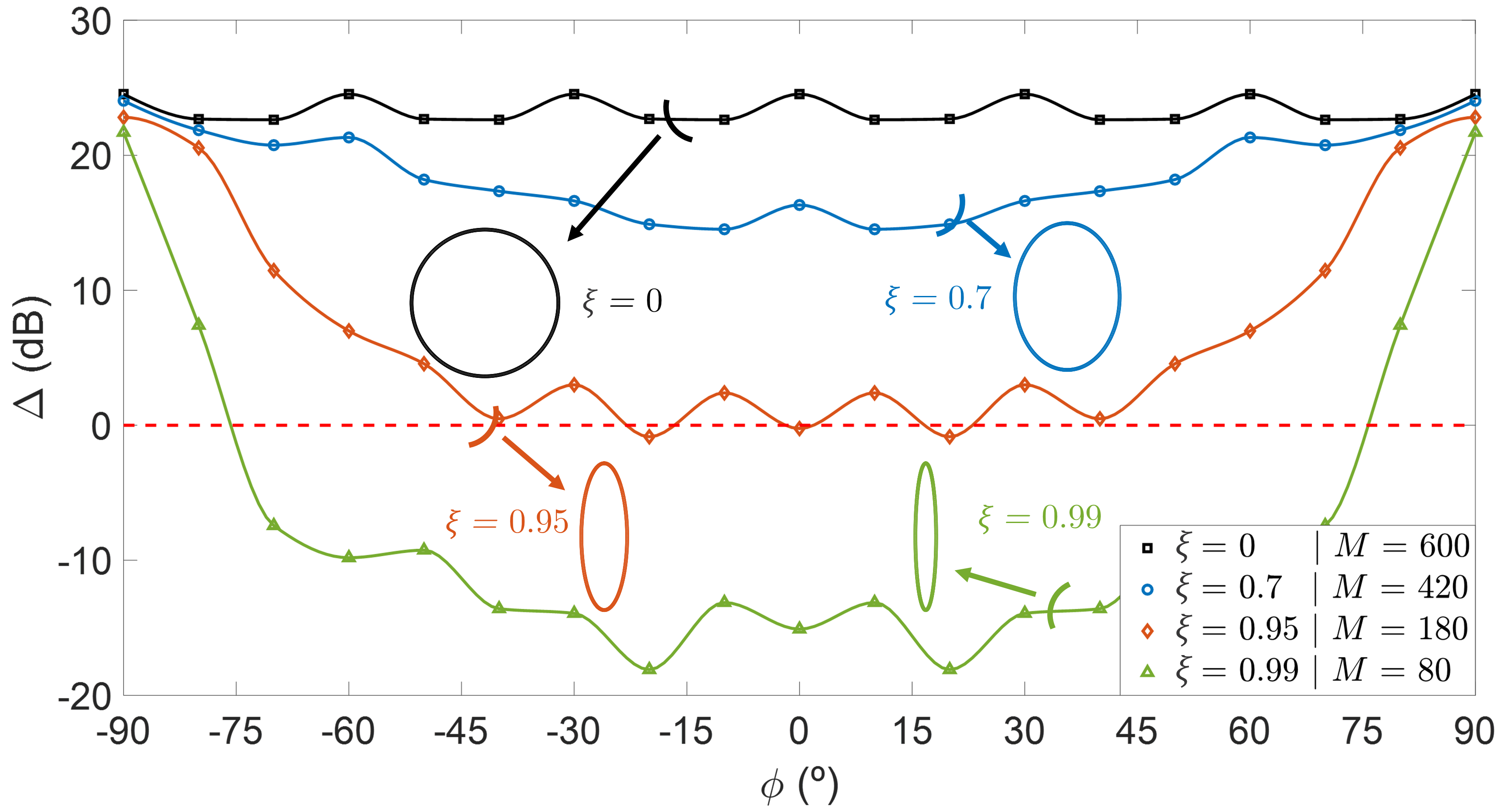}
	}
	\subfigure[]{\includegraphics[width= 1\columnwidth]{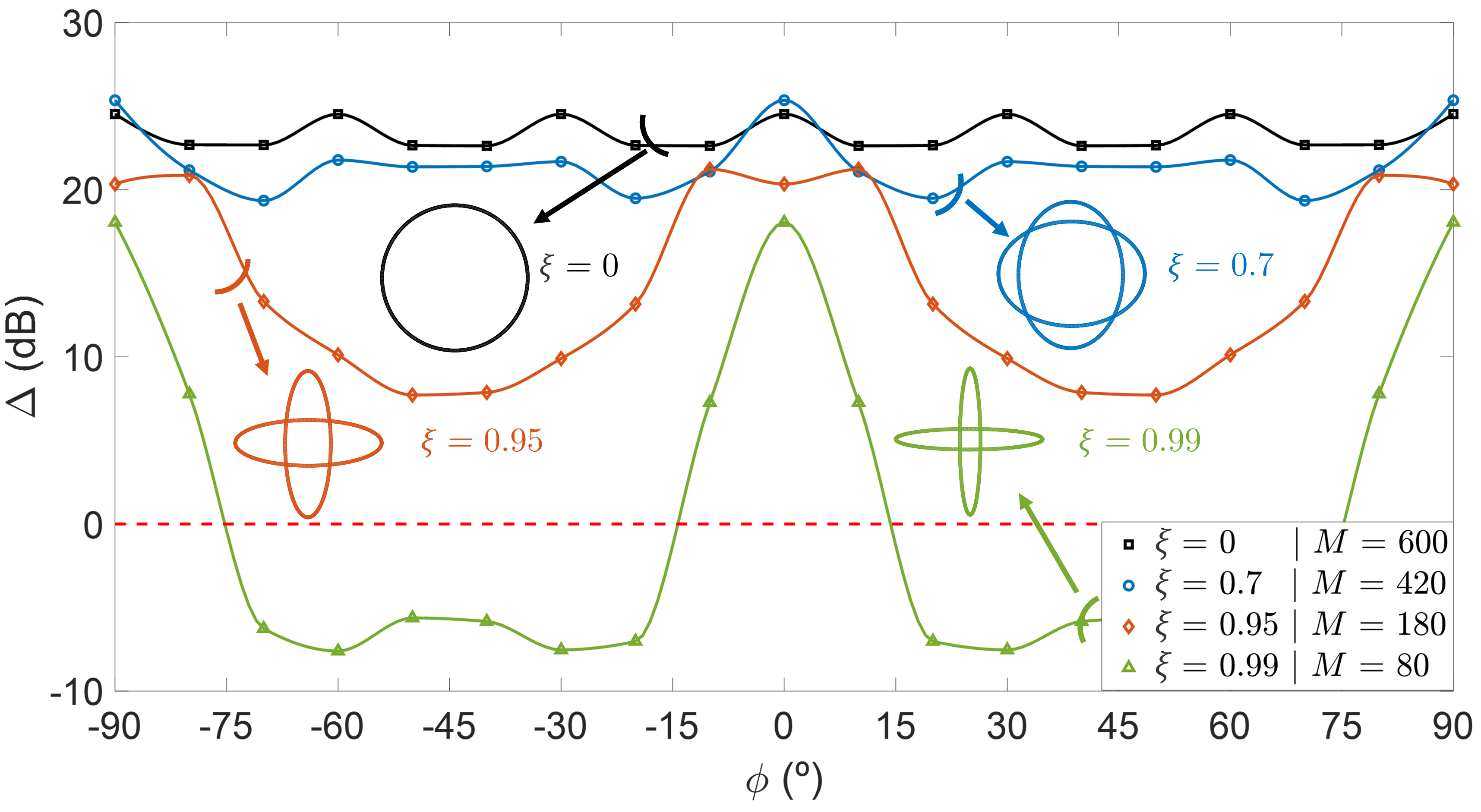}
	}
	\caption{ Metric $\Delta$ (dB) for several incident azimuth angles $\phi_l$ and several eccentricities $\xi$. a) Single ellipse with $\alpha = 0\degree$, b) single ellipse with $\alpha = 90\degree$, and c) two concentric ellipses with $\alpha = 0\degree$ and $\alpha = 90\degree$. These geometries show angular selectivity for $\phi_l = 0\degree$ and $\phi_l = 90\degree$.} 
	\label{delta_elipses}
\end{figure}

Fig. \ref{delta_elipses}(a) shows the metric $\Delta$  for several azimuth angles $\phi_l$ and eccentricity values $\xi$. $\Delta$~is constant for the circular array ($\xi = 0$) through the whole azimuth range due to the constant directivity at any given azimuth angle. When the eccentricity increases, the ellipse tends to flatten on the semi-minor axis. Therefore, the directivity of the estimation increases in the direction of the semi-major axis. For $\xi = 0.7$, the estimation is correct in the whole range. For $\xi = 0.95$, $\Delta>0$ for azimuth values up to $\pm50\degree$, where the largest artifact begins to mislead the correct incident angle. The same behavior can be found for $\xi=0.99$, where the estimation is valid up to $\pm 15\degree$. The reasoning behind these results is that for very high eccentricities, the angular response of an elliptical array closely resembles that of a linear array in the direction of the semi-major axis. Therefore, when the incident wave is perpendicular to the semi-major axis of the elliptical array, ambiguity affects the estimation and, consequently, an artifact appears at $\phi_l + 180\degree$, since the near-zero curvature of the array only exploits half of the angular domain, i.e., $180\degree$. This evolution can be seen as the eccentricity increases and the array flattens out in Fig.~\ref{delta_elipses}(a). As previously explained in Fig.~3, $\Delta$ value is calculated by taking into account the highest artifact found in the DoA and ToA estimation since $\phi_l$ can take any value on the entire range, i.e., $\phi_l \in [0\degree, 360\degree)$. However, if there were information on a bounded $\phi_l$ range, it would be possible to discard artifacts outside that range, thus improving the estimation of $\Delta$ value if the highest artifact is not part of the considered interval. Note that the formulation derived in Sec. \ref{sec:Theory} is valid for circular, elliptical and linear arrays. Although it is not explicitly shown, due to the array symmetry, the estimation is similar in the range $\phi_l = [90\degree, 270\degree]$. Also notice that the larger the eccentricity, the lower the number of considered phase modes $M$, as detailed in Sec. \ref{sec:Theory}. The computation time, as shown in the complexity analysis (see Section II.A), is expected to be higher for elliptical arrays. In the circular array case, the DoA and ToA computation time is 7.39 s. For the elliptical arrays with eccentricities 0.7, 0.95 and 0.99, computation times are 98.11 s, 54.06 s and 13.62 s, respectively. These times are obtained as the average for 10 iterations of the method in a laptop with an i7-10750H processor and 16 GB RAM. \blue{Table I summarizes the main features of the different geometries. $t_B$ stands for the required time to numerically compute the Bessel functions for the filters $W_{m,p}(f)$ and $t_f$ is the required time to apply $W_{m,p}(f)$ to $H_{p,l}(f)$ and compute the 2-D FFT. $t_{t}$ is the sum of the two previous times. Despite the longer required time, elliptical arrays provide three degrees of freedom that generalize the circular case and allow the creation of pseudorandom patterns discussed in depth in Section III.D.}

\renewcommand{\arraystretch}{1.5}
\setlength{\tabcolsep}{4pt}

\begin{table*}
    \centering
    \caption { \blue{Computational complexity, computation time, degrees of freedom and $\Delta$ for several geometries} } \label{tabla_comparativa} 
    \resizebox{\textwidth}{!}{
    
    \begin{tabular}{c|c|c|c|c|c|c|c|c|c}
    \hline \hline
    \multicolumn{1}{c|}{\blue{Geometry}} & \multicolumn{1}{c|}{\makecell{\blue{Computational}\\\blue{complexity}}} & \multicolumn{1}{c|}{\blue{Parameters}} & \multicolumn{1}{c|}{\blue{$t_{B}(\textrm{s})$}} & \multicolumn{1}{c|}{\blue{$t_{f}(\textrm{s})$}} & \multicolumn{1}{c|}{\blue{$t_{t} (\textrm{s})$}} & \multicolumn{1}{c|}{\blue{$\Delta (\textrm{dB})$}} & \multicolumn{1}{c|}{\makecell{\blue{Degrees}\\\blue{of freedom}}} & \multicolumn{1}{c|}{\makecell{\blue{Pseudorandom}\\\blue{patterns}}} & \multicolumn{1}{c}{\makecell{\blue{Angular}\\\blue{selectivity}}} \\
    \hline \hline
     
     
    \hline \blue{Circular} &  \blue{$\mathcal{O}(M\,\Psi)$} & \blue{$r = 0.5 \textrm{ m}$} & \blue{$1.96$} & \blue{$5.43$} & \blue{$7.39$} & \blue{$24.53$} & \blue{$1$} & \blue{\xmark} & \blue{\xmark} \\
    \hline
      & $ $ & \blue{$a = 0.5 \textrm{ m}$, $\alpha = 0\degree$, $\xi = 0.7$} & \blue{$93.68$} & \blue{$4.43$} & \blue{$98.11$} & \blue{$24.07$} & $ $ & $ $ & $ $ \\
    \blue{Elliptical} & \blue{$\mathcal{O}(PM\,\Psi)$} & \blue{$a = 0.5 \textrm{ m}$, $\alpha = 0\degree$, $\xi = 0.95$} & \blue{$52.58$} & \blue{$1.48$} & \blue{$54.06$} & \blue{$22.81$} & \blue{$3$} & \blue{\cmark} & \blue{\cmark}\\
     & $ $ & \blue{$a = 0.5 \textrm{ m}$, $\alpha = 0\degree$, $\xi = 0.99$} & \blue{$13.04$} & \blue{$0.58$} & \blue{$13.62$} & \blue{$21.70$} & $ $ & $ $ & $ $\\
    \hline \hline
    \end{tabular}

    }
    
\end{table*}

\begin{figure}[t]
	\centering
	\includegraphics[width= 1\columnwidth]{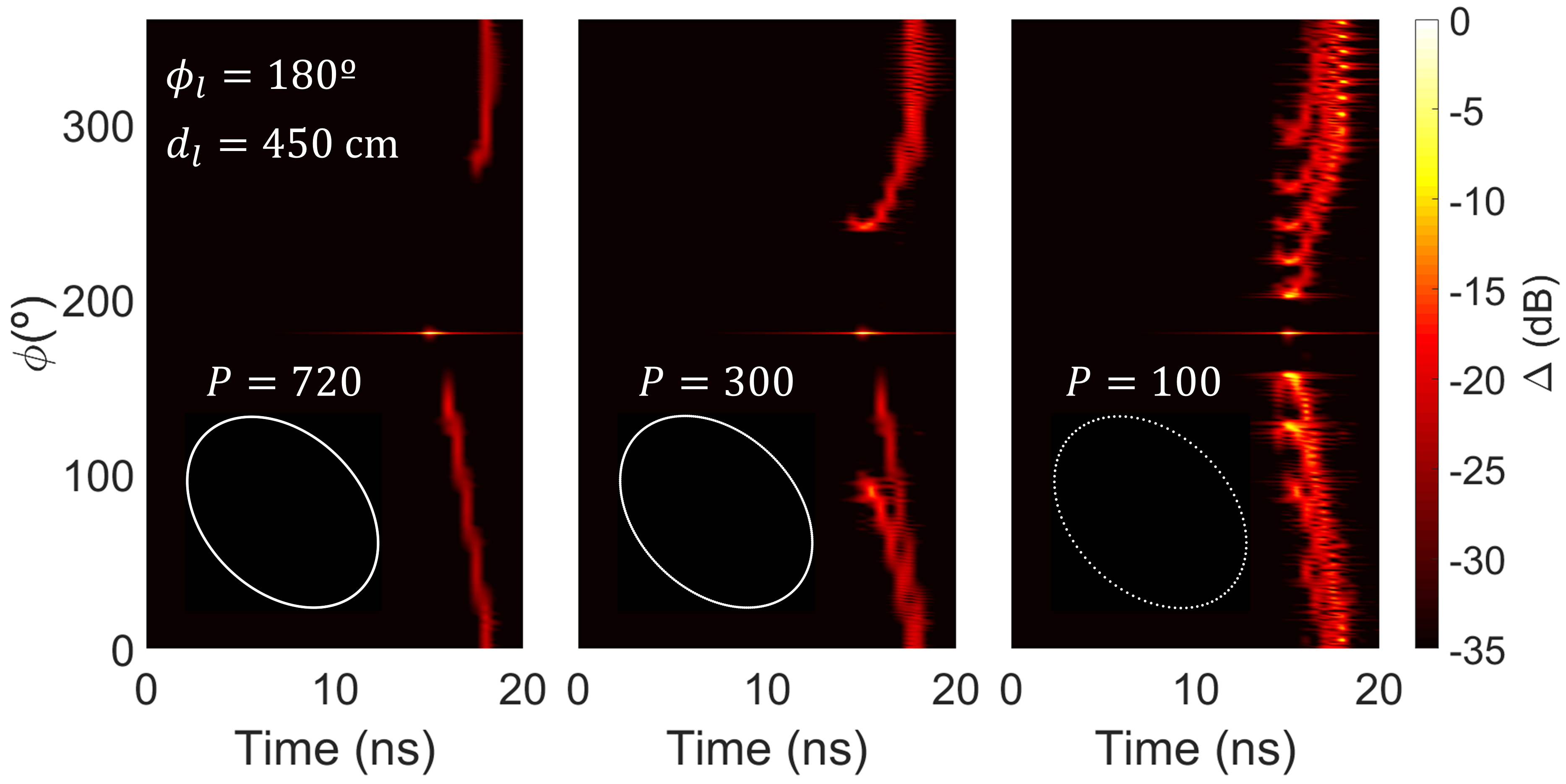}
	\caption{\blue{DoA and ToA estimation for an incident wave at $\phi_{l} = 180\degree$ and $\tau_{l} = 15 \text{ ns } (d_{l} = 450 \text{ cm})$ when varying the number of sensors $P$. Non-compliance with the spatial Nyquist theorem (cases $P=300, 100$) leads to the appearance of new artifacts and an undesired increase in their contribution to the joint DoA and ToA estimation.}} 
	\label{reduction}
\end{figure}

At the beginning of this Section, three degrees of freedom related to ellipses were discussed: semi-major axis, eccentricity and rotation angle. As Fig.~\ref{delta_elipses}(a) has shown, highly flattened ellipses are angularly selective along direction of the semi-major axis. In order to tune the angular response, Fig.~\ref{delta_elipses}(b) analyses the effect of a rotation angle $\alpha$ in the ellipse geometry. Particularly, it includes a rotation angle $\alpha=90\degree$ compared to the geometry of Fig.~\ref{delta_elipses}(a). The direct consequence is the $90\degree$ shift on the azimuth angle estimation. Now, the semi-major axis is located in the vertical axis, providing excellent estimation for very flattened ellipses when the incident angle is nearby $\pm 90\degree$. However, incident angles in the horizontal axis ($0\degree$ and $180\degree$) get poor estimations if eccentricities are large, contrary to what is observed in Fig.~\ref{delta_elipses}(a). \blue{Additionally, as it is previously stated, the number of sensors can be reduced at the expense of degrading the estimation performance. In order to show the effect of this reduction, Fig.~\ref{reduction} shows the joint DoA and ToA estimation for an incident wave at $\tau_l = 15$~ns $(d_l = 450$ cm$)$ and $\phi_l = 180\degree$ in an elliptical array with $a = 0.5$ m, $\alpha = 135\degree$ and $\xi = 0.7$. Three different numbers of sensors $P = 720, 300$ and $100$ are considered. While for $P = 720$, the maximum spacing between sensors remains below $\lambda/2$, for $P = 300$ and $100$, these distances becomes $1.048\lambda$ and $3.143\lambda$. This causes the artifacts to move from 19.34 dB below the estimation to 13.73~dB and 5.99~dB respectively. Thus, although it is possible to perform the estimation even for adjacent sensor distances larger than $\lambda/2$, it is advisable to comply with the spatial Nyquist theorem to minimize the appearance of artifacts.}

In summary, the proposed formulation performs proper estimation of DoA and ToA for the whole azimuth range for eccentricity values up to 0.7. Above this value, the estimation shows a directional behavior in the direction where the semi-major axis is located. By rotating the ellipses, the angular response can be tuned. This can be easily achieved in virtual arrays. Logically, mechanical or electrical reconfiguration in a real-world implementation always adds an extra level of technical complexity that is beyond the features of the method and the scope of the present work. Thus, the angular selectivity of the elliptical geometries could be of potential application for the selection (suppression) of signals-of-interest (signals-not-of-interest) in smart wireless environments.

\subsection{Concentric Elliptical Arrays}

Sec.~\ref{sec:Theory}.C introduced the analysis of concentric elliptical arrays. Since previous subsection has shown the eccentricity effect on the estimation, we can take advantage of the concentric arrays in order to improve the estimation. By combining $H_{m,l}(f)$ from the ellipses shown in Figs. \ref{delta_elipses}(a) and \ref{delta_elipses}(b) [see eq. (\ref{Hml_psi2})], a new ellipse arrangement can be formed. Fig.~\ref{delta_elipses}(c) presents the metric $\Delta$ for the joint DoA and ToA estimation when two concentric elliptical arrays are considered. For high eccentricities, an interesting behavior can be found. $\Delta$ is maximized for all those DoA coinciding with the semi-major axes of the concentric ellipses (i.e., $\phi_l = 0\degree$ and $\phi_l = 90º\degree$). A clear example can be seen for $\xi = 0.95$. In Figs. \ref{delta_elipses}(a) and \ref{delta_elipses}(b), $\Delta$ indicates good estimations for approximately an $80\degree$ azimuth range. However, in Fig.~\ref{delta_elipses}(c), $\Delta>8$ dB in the whole range. Therefore, concentric ellipses with different rotation angles can provide different $\Delta$ patterns. This fact opens up the possibility of implementing pseudorandom grids contained within elliptical geometries, which will be discussed in later sections.

\begin{figure}[t]
	\centering
	\includegraphics[width= 1\columnwidth]{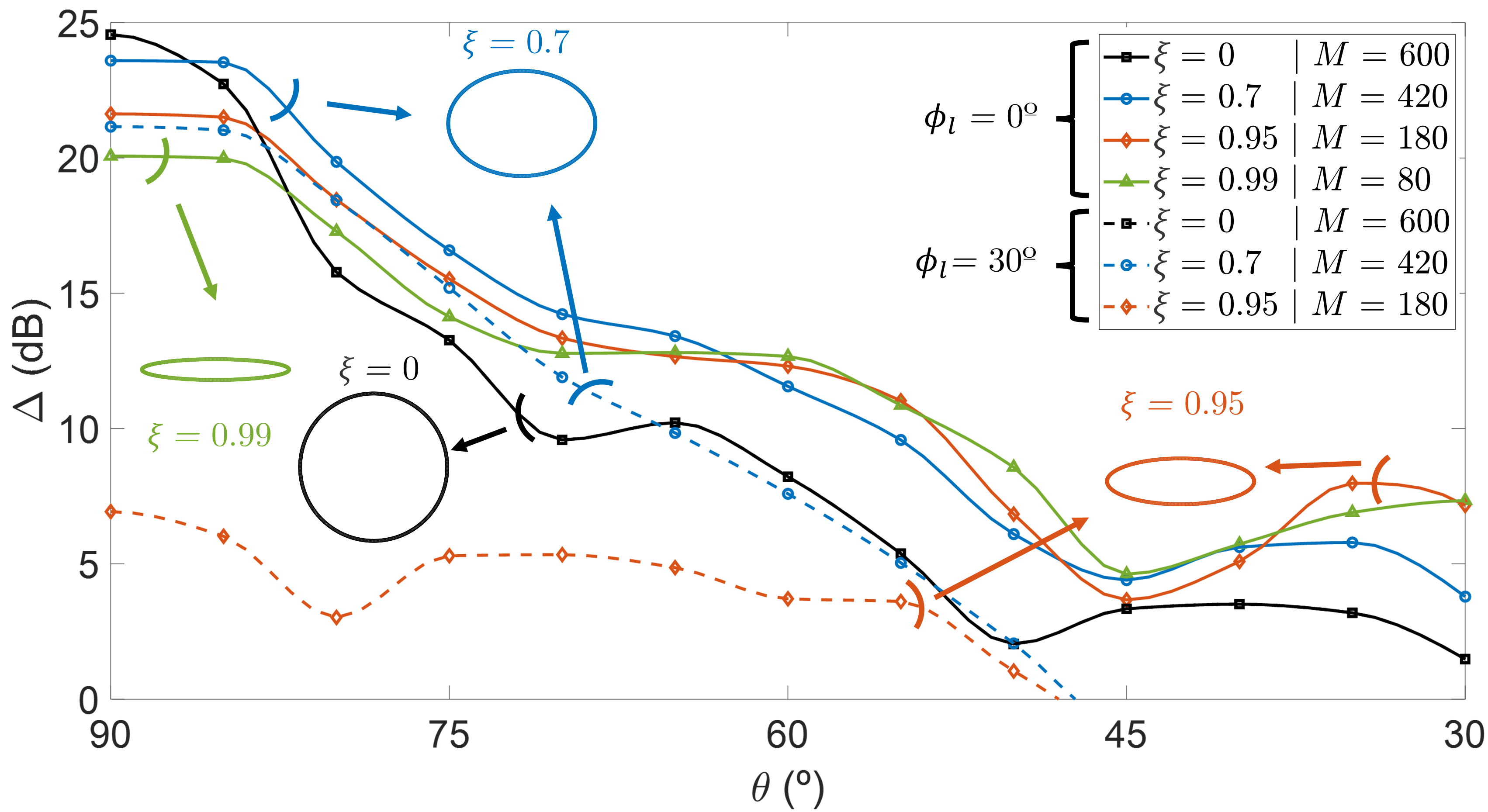}
	\caption{Metric $\Delta$ (dB) for several elevation angles $\theta_l$ when $\phi_l = 0\degree$ (solid line) and $\phi_l = 30\degree$ (dashed line).}
	\label{fig_elevacion}
\end{figure}

\subsection{Technique performance for $\theta_l \neq 90^\mathrm{o}$}

Up to this point, simulations have been carried out for a fixed elevation angle of $\theta_l = 90^\degree$.  In \cite{Zhang_2017}, it was demonstrated that the filter from eq. \eqref{Wm2} provides accurate estimations for $\theta_l \neq 90^\degree$. With the purpose of validating the estimation for different elevation angles, Fig.~\ref{fig_elevacion} shows the metric $\Delta$ for different values of $\theta_l$ for two different incident azimuth angles, $\phi_l = 0\degree$ and $\phi_l = 30\degree$. The configuration parameters are the same as those shown in Sec. III.A. For the case $\phi_l = 0\degree$, although the analytical framework discussed in Sec. II was originally derived under the assumption of $\theta_l=90\degree$, it can be observed in Fig.~\ref{fig_elevacion} that the accuracy in DoA estimation is still satisfactory in a wide range of elevation angles. This is a remarkable feature, as elevation/tilt angles typically vary between 90 and 75 degrees in real deployments \cite{tilt_1, tilt_2}. In this range, $\Delta>13$ dB for all the considered eccentricities. Concerning ToA, it is estimated as 30 ns for $\theta_l = 90\degree$, matching $\tau_l$. When $\theta_l$ moves toward $30\degree$, ToA estimation suffers a slight variation of $+0.5$ ns. This effect was also noticed for circular arrays in \cite{Fan_2019}. Finally, it is worth noting that elliptical arrays outperform DoA and ToA estimation compared to circular arrays for several elevation angles when $\phi_l = 0\degree$. This is due to the fact that when the angle of incidence $\phi_l$ coincides with the semi-major axis, the elliptical shape of the array preserves the directivity of the array better for $\theta_l \neq 90\degree$ compared to the circular shape. For the case $\phi_l = 30\degree$, $\Delta$ with $\theta_l = 90\degree$ for high eccentricities is lower as previously depicted in Fig. \ref{delta_elipses}(a) due to the directivity attribute. Similar to the case $\phi_l = 0\degree$, $\Delta$ decreases as the wave incident plane separates from the plane where the elliptical array lies.

\subsection{Pseudorandom Grids}

\begin{figure}[t]
	\centering
	\includegraphics[width= 1\columnwidth]{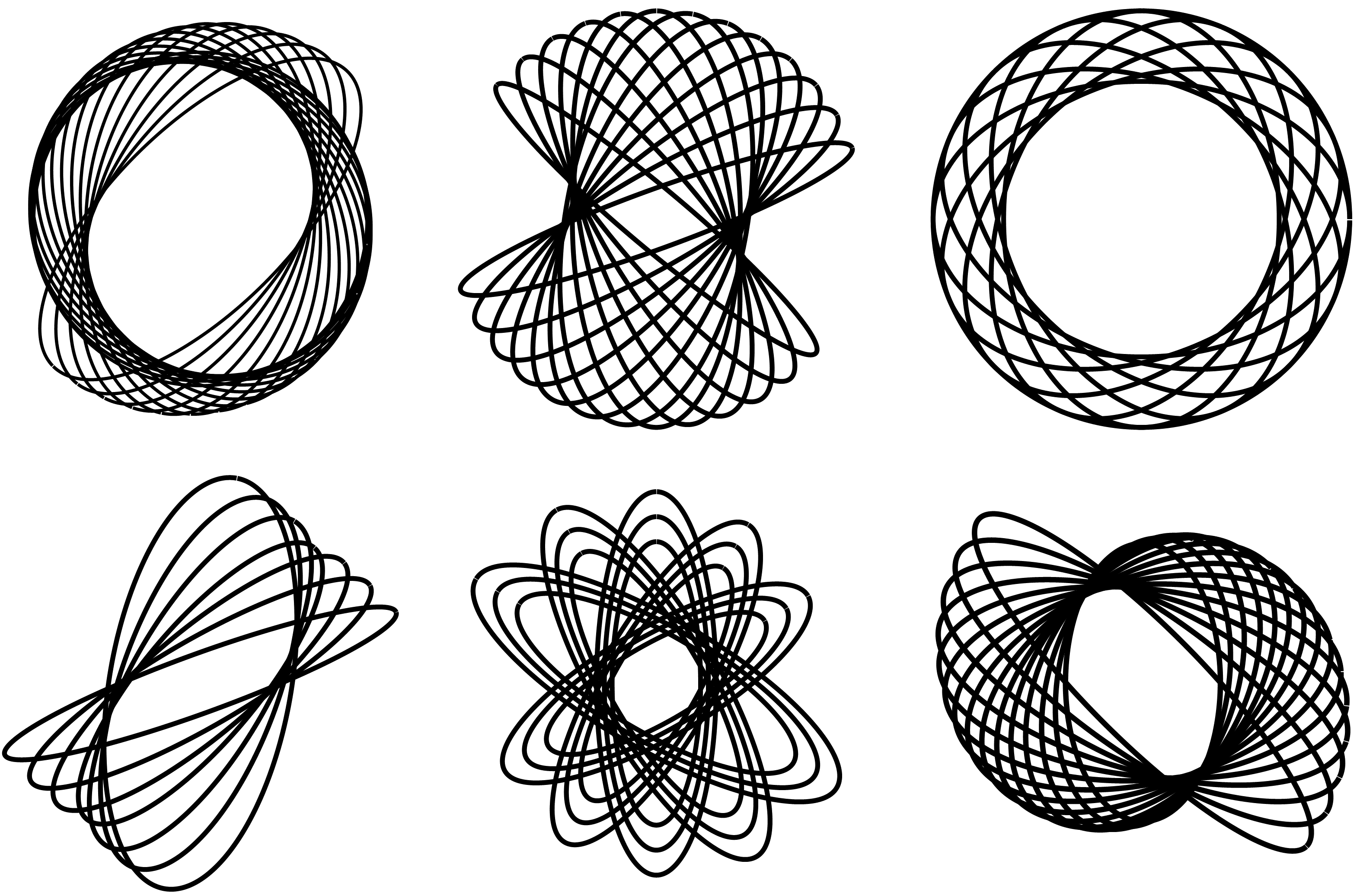}
	\caption{Generic geometries generated from the superposition of concentric ellipses. Each ellipse is characterized by semi-major axis $a$, eccentricity $\xi$ and angle of rotation $\alpha$. The combination of several ellipses results in pseudorandom grids.} 
	\label{esquema_elipses_superpuestas}
\end{figure}

Previous subsections have shown that elliptical arrays present directive behaviour for DoA estimation. Fig. \ref{esquema_elipses_superpuestas} shows some examples of generic geometries formed through the superposition of concentric ellipses. In contrast to concentric circular arrays where only concentric ring shapes can be obtained \cite{phase_mode2007}, concentric ellipses provide a wide range of possible sensor arrangements. Hence, some of these geometries may be approximated as pseudorandom grids and the joint DoA and ToA estimation could be performed with sensors located in pseudorandom positions. In real deployments, the use of circular arrays may not be feasible due to space limitations. Conversely, elliptical arrays take up a smaller area and can adapt better to the geometry of any structure. Additionally, we can take advantage of the directivity property in the direction of the semi-major axis to steer the elliptical arrays in a specific range of angles.

\begin{figure}[t]
	\centering
	\includegraphics[width= 1\columnwidth]{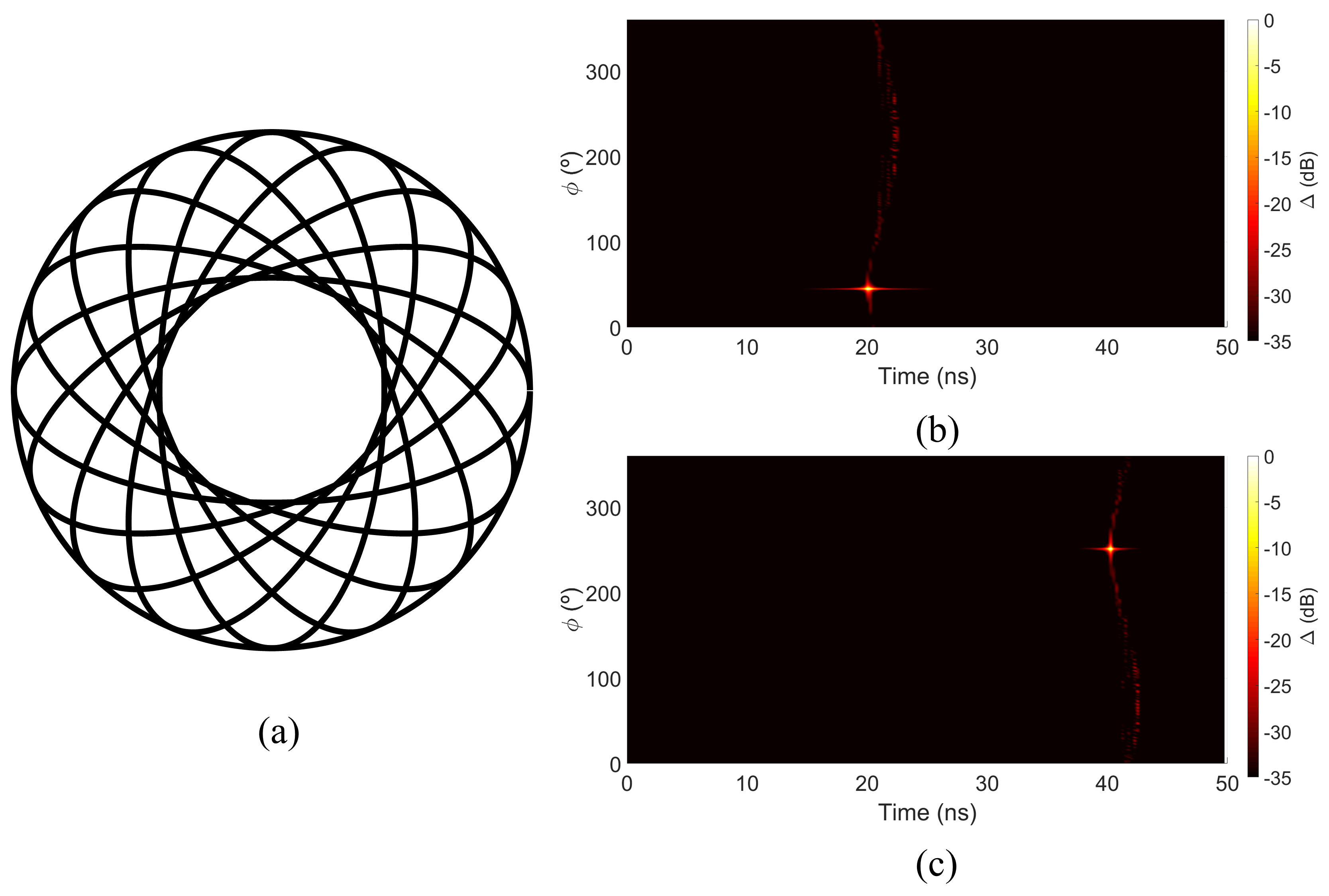}
	\caption{(a) Superposition of nine concentric elliptical arrays.  DoA and ToA estimation when the incident wave is located at (b) $\phi_l = 45\degree$ and $\tau_l = 20$~ns, and (c) $\phi_l = 250\degree$ and $\tau_l = 40$~ns.} 
	\label{superposicion_estudio}
\end{figure}

As a study case, Fig. \ref{superposicion_estudio}(a) shows a superposition of concentric ellipses to form a pseudorandom grid. Throughout this subsection, it will be studied in depth to analyze its performance compared to single and concentric circular arrays. This geometry consists of 8 ellipses with semi-major axis $a = 34.5$ cm, eccentricity $\xi = 0.9$ and rotation angle $\alpha = 22.5\degree$ for consecutive ellipses, plus an outer circle of $a = 34.5$ cm and $\xi = 0$. In this case, the frequency band goes from 39.5~GHz to 43.5~GHz (n259 band and $B$~=~4~GHz) \cite{3GPP_TS}. 720~sensors are considered per ellipse, which gives a maximum separation between consecutive sensors of 3 mm (0.437$\lambda$ at $f = 43.5 \textrm{ GHz})$. The number of considered frequency samples is $K=200$ and the number of phase modes is fixed to $M = 250$.

By applying the formulation for concentric elliptical arrays, Figs.~\ref{superposicion_estudio}(b) and \ref{superposicion_estudio}(c) present two cases of joint estimation for this geometry. Particularly, Fig.~\ref{superposicion_estudio}(b) shows the estimation of the angular-delay domain for an incident wave at $\phi_l = 45\degree$ and $\tau_l = \nolinebreak20$ ns. Fig.~\ref{superposicion_estudio}(c) illustrates the same domain for  $\phi_l =\nolinebreak250\degree$ and $\tau_l = 40$ ns. In both cases, the estimation is clearly maximized around ToA and DoA. In addition to these cases, the geometry has been tested for incident waves in the entire range of $\phi_l$ and $\tau_l$, showing excellent results with artifacts below 20 dB in the worst case.

\begin{figure}[t]
	\centering
	\subfigure[]{\includegraphics[width= 1\columnwidth]{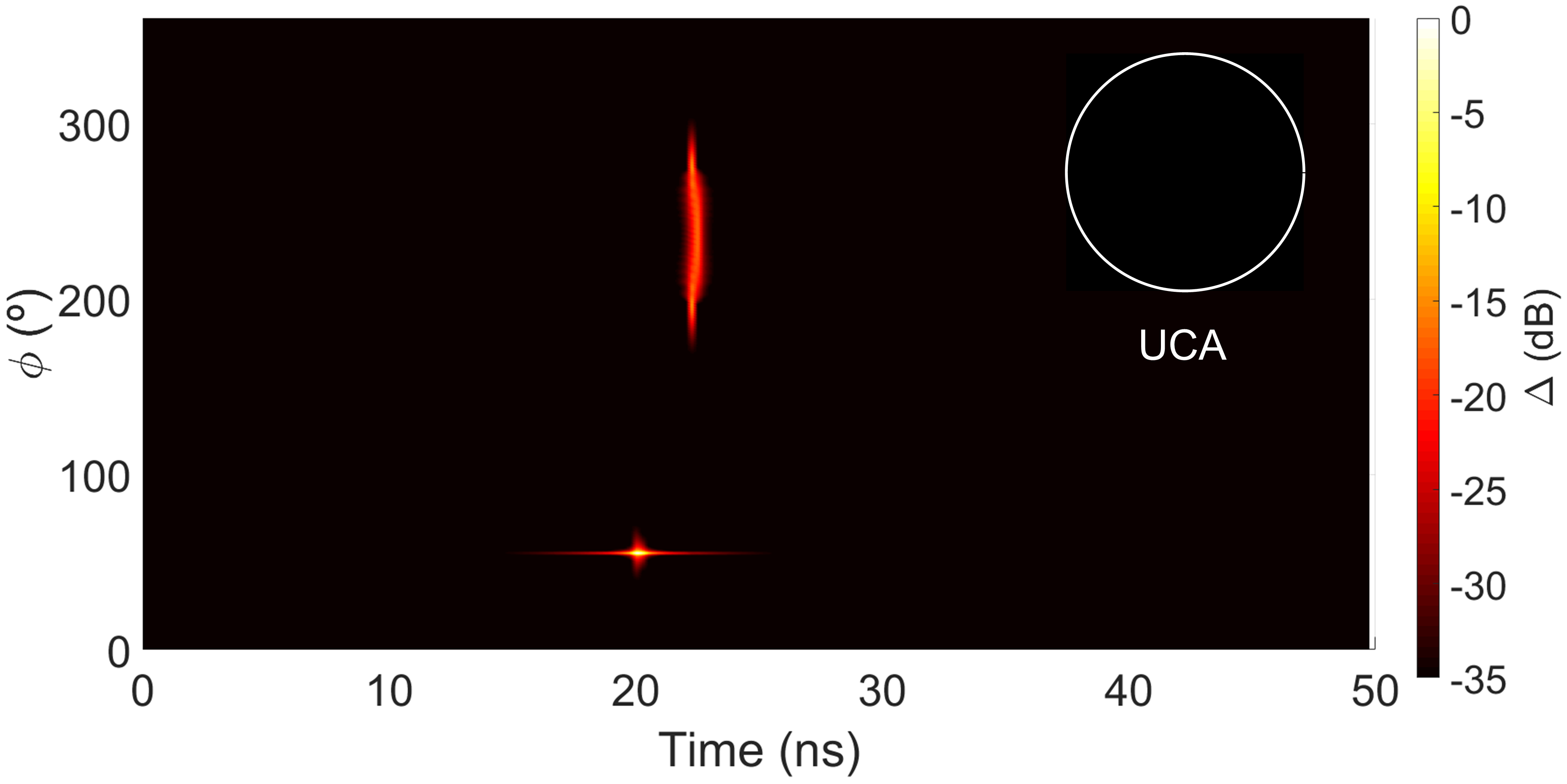}
	}
	\subfigure[]{\includegraphics[width= 1\columnwidth]{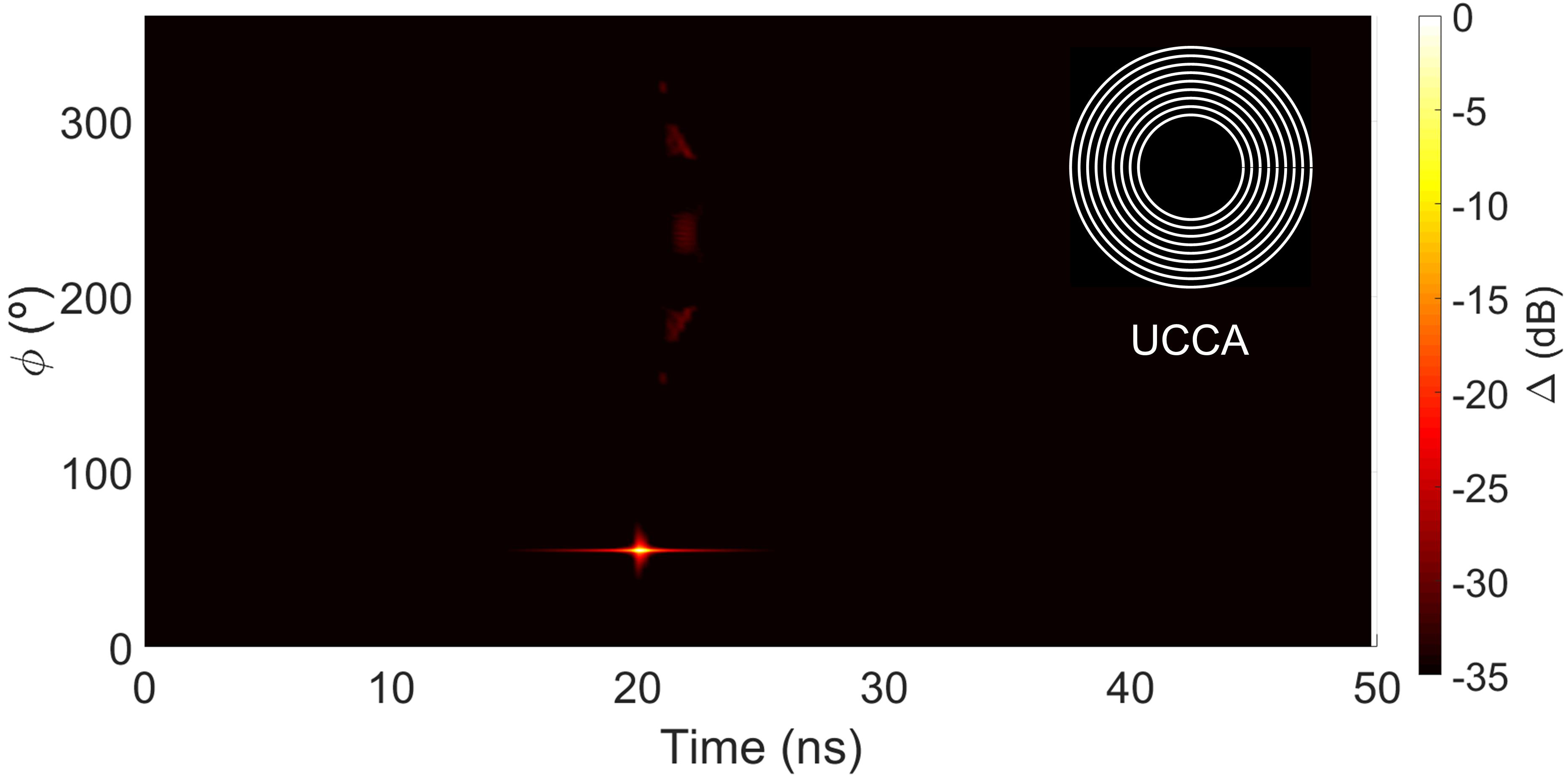}
	}
	\subfigure[]{\includegraphics[width= 1\columnwidth]{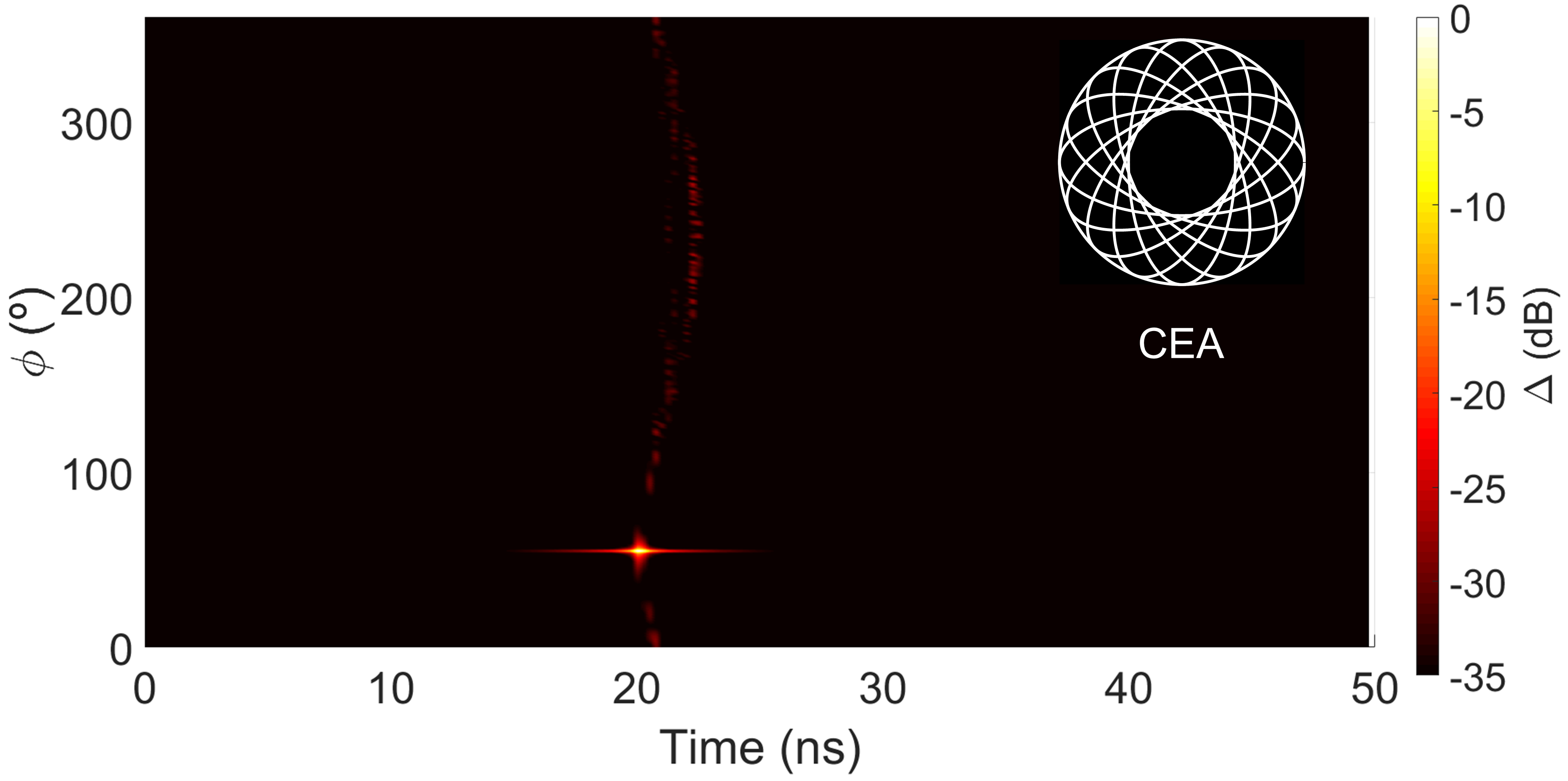}
	}
	\caption{DoA and ToA estimation for an incident wave at $\phi_l = 55º\degree$ and $\tau_l = 20$ ns for different geometries: (a) Uniform Circular Array, (b) Uniform Concentric Circular Array, and (c) Concentric Elliptical Array.} 
	\label{fig_UCA_UCCA_CEA}
\end{figure}

In order to compare the performance of concentric elliptical arrays with other array arrangements, Fig.~\ref{fig_UCA_UCCA_CEA} shows the angular-delay estimation for three different arrangements. Fig.~\ref{fig_UCA_UCCA_CEA}(a) is obtained from a uniform circular array (UCA) with radius $r = 34.5$ cm and $P = 720$. Fig.~\ref{fig_UCA_UCCA_CEA}(b) represents the estimation for a nine ring uniform concentric circular array (UCCA) whose outer circle is equal to the one shown in Fig.~\ref{fig_UCA_UCCA_CEA}(a). The inner circle has radius $r = 15$ cm and all nine rings are equidistant. Finally, Fig.~\ref{fig_UCA_UCCA_CEA}(c) uses the geometry presented in Fig.~\ref{superposicion_estudio}(a), i.e., a concentric elliptical array (CEA). 
Some conclusions can be extracted by looking at Fig.~\ref{fig_UCA_UCCA_CEA}. First, the figure illustrates that the proposed method works as a generalization of former approaches, being able to deal with circular and elliptical geometries at the same time. This leads to the realization of elliptical-based pseudorandom mesh grids that can be used to improve the joint DoA and ToA estimation. In that sense, the level of artifacts (sidelobes) has been reduced more than 10 dB when considering concentric arrays instead of a single circular array, the UCCA and CEA outperforming the UCA array due to a better mapping of the spatial region. Naturally, the improvement in the estimation comes at a price. Computational complexity increases when a greater number of arrays is considered. Thus, there exists a trade-off between performance and computational complexity. Scenarios where the impact of the artifacts in the joint estimation should be minimized benefit for the inclusion of a greater number of concentric (elliptical) arrays. Conversely, scenarios where the processing time should be minimized benefit from placing a fewer number of arrays.


\begin{figure}[t]
	\centering
	\subfigure[]{\includegraphics[width= 1\columnwidth]{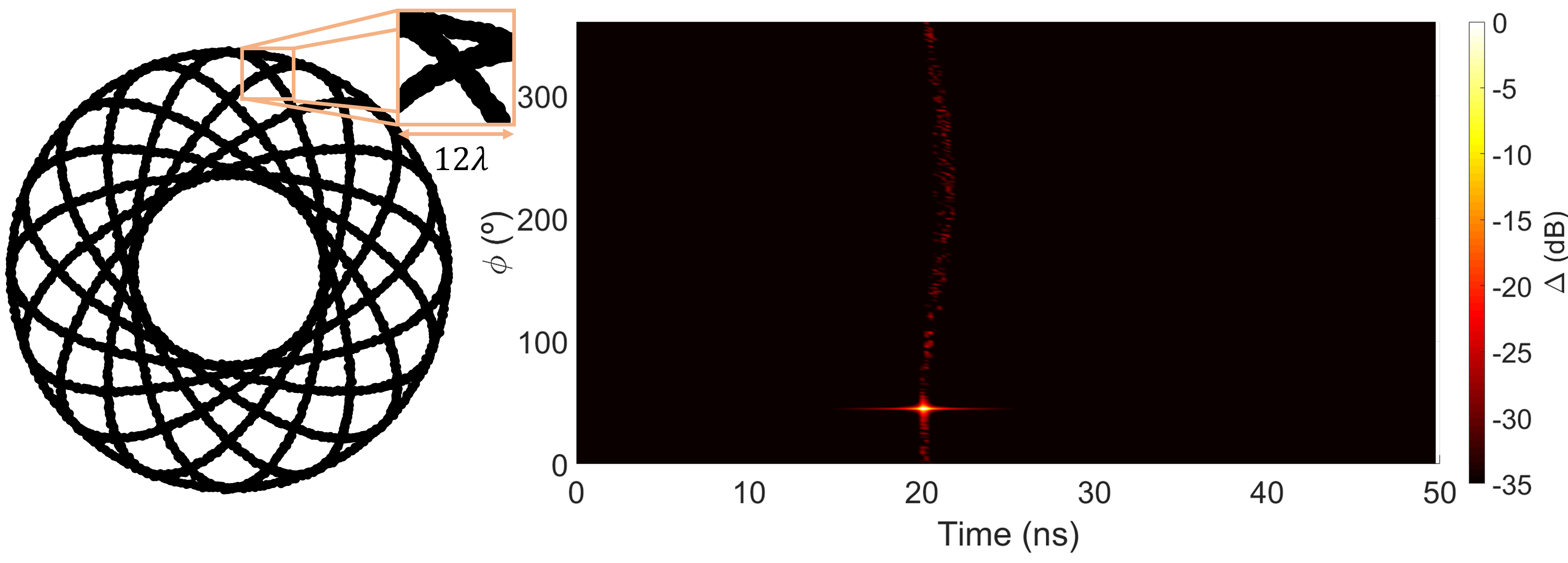}
	}
	\subfigure[]{\includegraphics[width= 1\columnwidth]{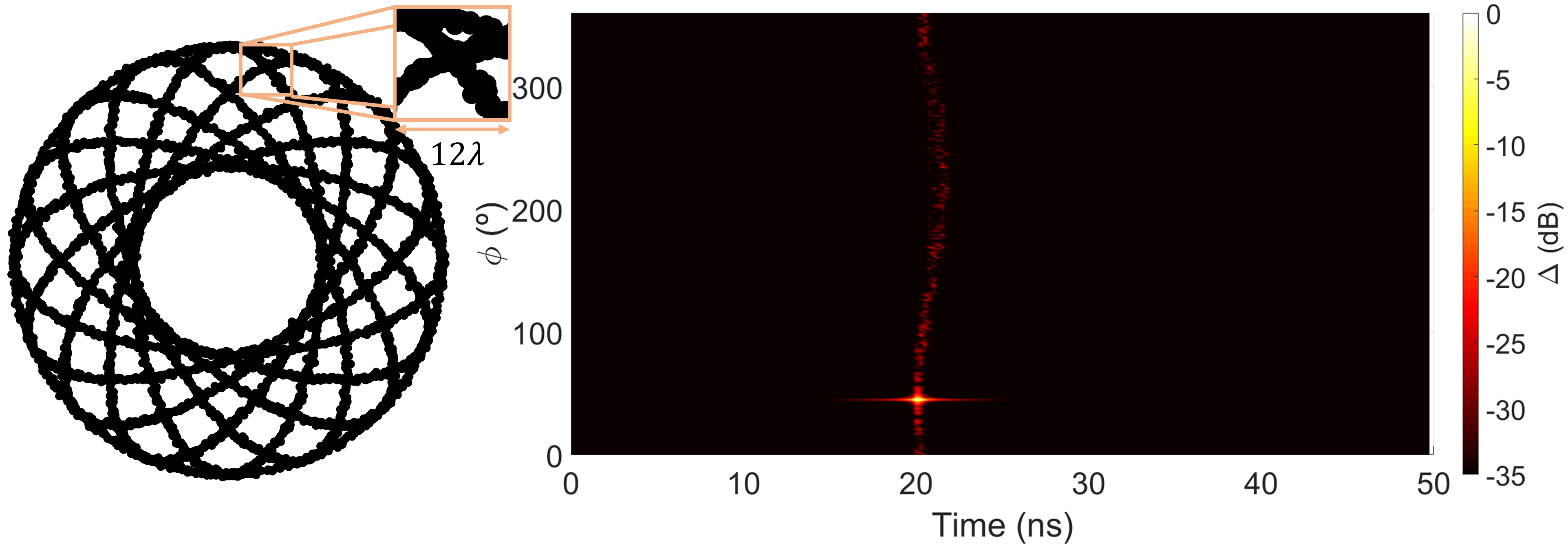}
	}
	\subfigure[]{\includegraphics[width= 1\columnwidth]{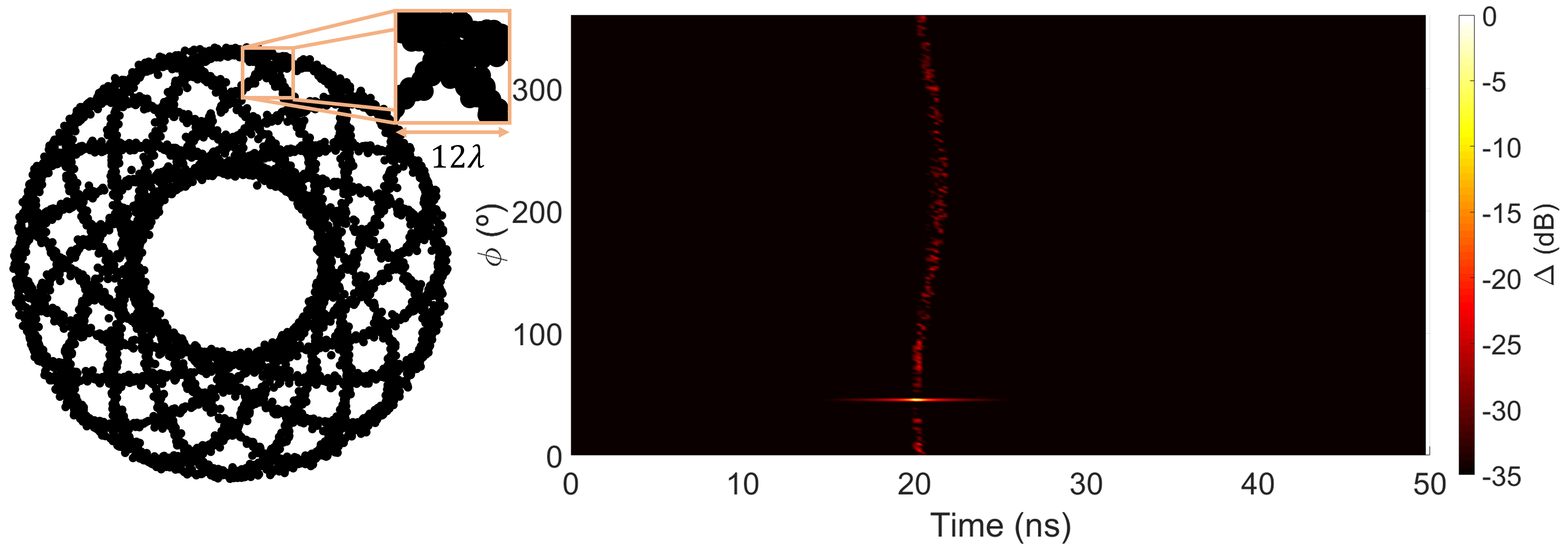}
	}
	\subfigure[]{\includegraphics[width= 1\columnwidth]{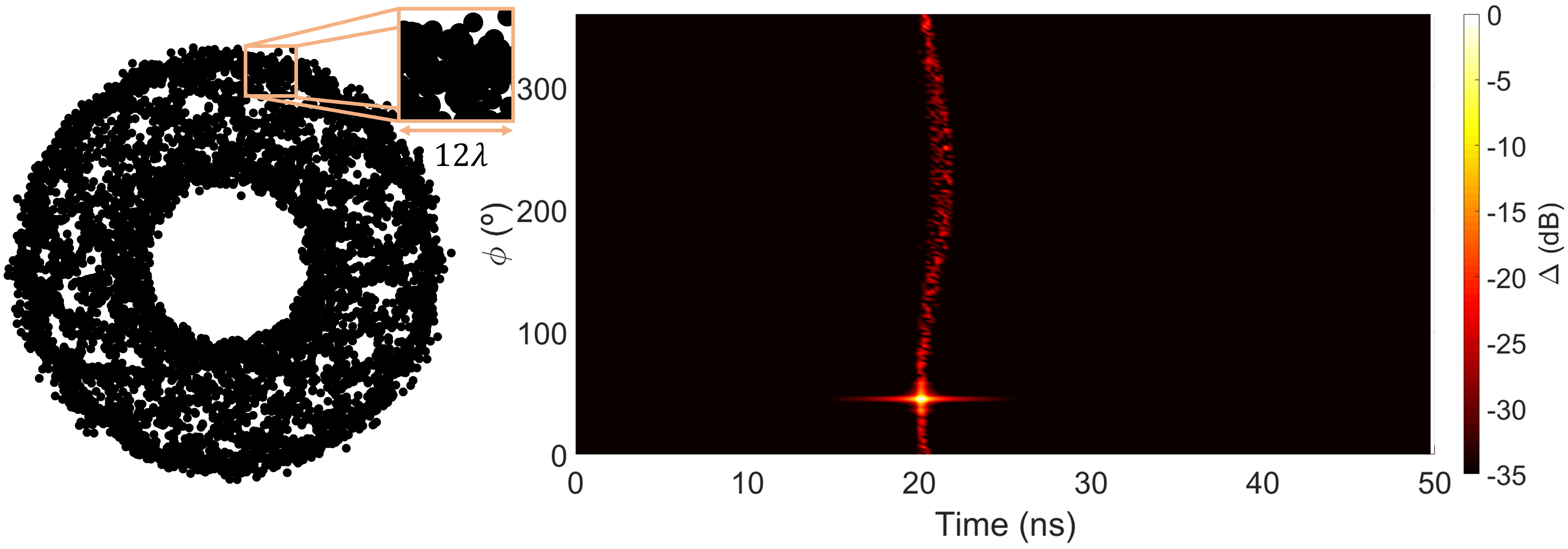}
	}
	\caption{DoA and ToA estimation when the incident wave reaches the CEA at $\phi_l = 45\degree$ and $\tau_l = 20$ ns. Uncertainty in sensor placement is considered, with respect to the center position \{$x_p, y_p$\},  by means of the standard deviation $\sigma$. Cases:  (a)~$\sigma = 0.5\lambda$ (3.45 mm), (b)~$\sigma = \lambda$ (6.89 mm), (c)~$\sigma = 2\lambda$ (13.78 mm) and (d)~$\sigma = 5\lambda$ (34.46 mm).}
	\label{fig_noise_position}
\end{figure}

 Previously, the degrees of freedom of the position of the sensors in an ellipse have been discussed. Actually, it is possible to go one step further. The theoretical framework developed in Sec. II has demonstrated that non constant $r_p$ values are feasible for the joint estimation. In order to achieve a real pseudorandom grid, an independent and uncorrelated bivariate normal distribution $\mathcal{N}_{2}(0,\,\sigma^{2})$, with $\sigma_x = \sigma_y = \sigma/\sqrt2$, is included in the sensor positions $x_{p}$ and $y_{p}$ [eq.~\eqref{eq_rotada}]. This distribution adds noise in the sensor position, which randomizes the grid. Figs.~\ref{fig_noise_position}(a)-(d) shows the angular-delay domain estimation for arrangements with different $\sigma$ values. Simulation parameters are equal to those ones shown for Fig.~\ref{superposicion_estudio} with incident wave at $\phi_l = 45\degree$ and $\tau_l = 20$~ns. Standard deviation $\sigma$ is chosen in terms of wavelength and it ranges from $0.5\lambda$ to $5\lambda$. The maximum artifacts from Fig. \ref{fig_noise_position}(a) to Fig. \ref{fig_noise_position}(d) are found to be 23.9~dB, 23.5~dB, 21.4~dB and 18.2~dB below the correct estimation, respectively. Although artifacts increase as the position noise does, they are still low enough to correctly detect the DoA and ToA. A remarkable result is that shown in Fig. \ref{fig_noise_position}(d), where the elevated value $\sigma$ directly hides the elliptical shape of the CEA. Even in this case, the joint estimation is remarkably good. For a dense enough sensor arrangement, most of the sensors satisfy the spatial Nyquist theorem. If, on average, the spatial Nyquist theorem is fulfilled, a good estimation is expected. Therefore, it can be concluded that, if a dense pseudorandom mesh is approximated by a set of concentric ellipses, the joint estimation of DoA and ToA can be properly performed.

\begin{figure}[t]
	\centering
	\includegraphics[width= 1\columnwidth]{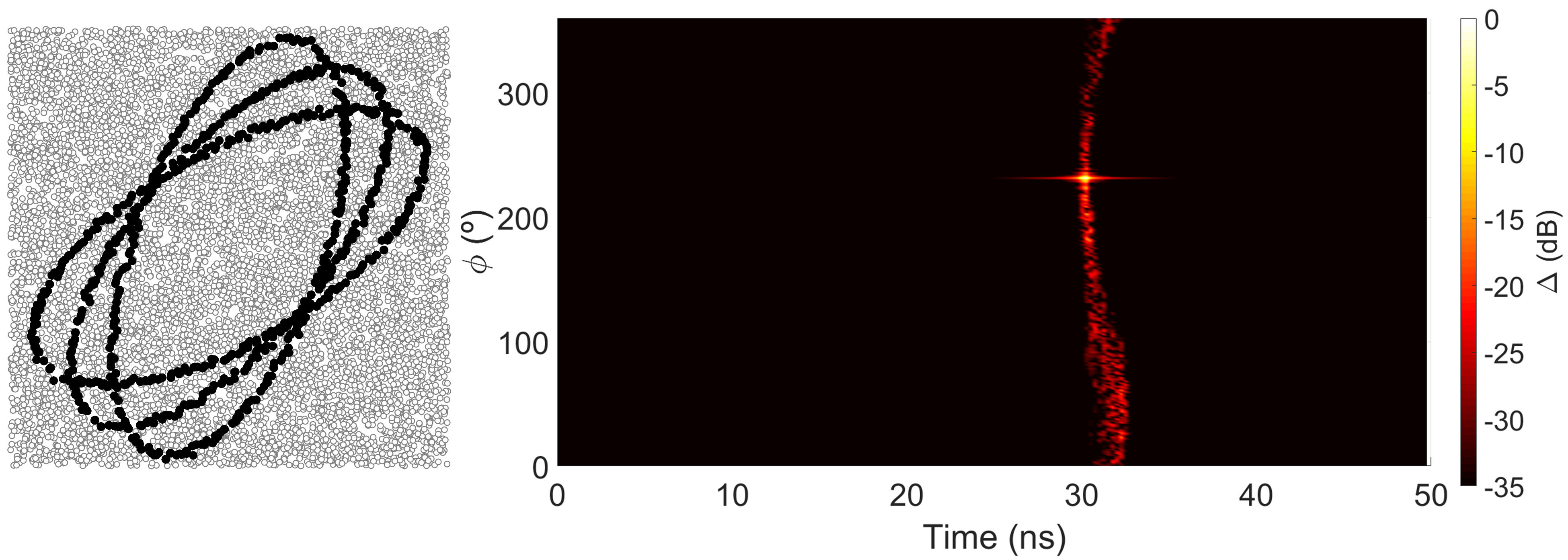}
	\caption{DoA and ToA estimation for an incident wave at $\phi_l = 230º\degree$ and $\tau_l = 30$ ns for a sensor distribution based on a bivariate uniform random distribution (gray dots). Three concentric elliptical arrays are approximated from the randomly arranged sensors (black dots).}
	\label{ROUND1_random}
\end{figure}

As a last proof of concept, the estimation of DoA and ToA from a completely random sensor distribution approximated by a set of concentric ellipses is proposed. For the sensor distribution, 10000 sensors are placed in a square based on a bivariate uniform distribution $\mathcal{U}_{\textrm{2 }}(-0.345, 0.345)$. As an example, this distribution expects to be approximated by three concentric ellipses with $P = 720$, $a = 34.5$ cm, $\xi = 0.9$ and $\alpha = [30\degree,\, 50\degree,\, 90\degree]$. Frequency band goes from 39.5 GHz to 43.5 GHz, $K = 200$ and $M = 250$. For this approximation, those sensors of the random distribution whose distance to the theoretical sensor in the three ellipses is minimum are chosen. Fig.~\ref{ROUND1_random} shows the random sensor distribution (gray dots) and the sensors chosen to form the concentric elliptical array (black dots). This figure also presents the DoA and ToA estimation for an incident wave at $\phi_l = 230\degree$ and $\tau_l = 30$ ns, with the largest artifact appearing 13.45 dB below the real path. Thus, the method is valid for elliptical arrays generated from random distributions. Note that the random grid can be used in order to generate any other geometry with several ellipses and several semi-major axis, eccentricity and rotation angle values. Finally, note that the pseudorandom and random grids shown in Figs.~\ref{fig_noise_position} and \ref{ROUND1_random} cannot be rigorously analyzed with previous approaches \cite{phase_mode2002, phase_mode2007, phase_mode2008, Zhang_2017, Fan_2019}, as sensors with different radius $r_p$ must be considered.

\section{\label{sec:Results} Measurements}

\begin{figure}[t]
	\centering
	\includegraphics[width= 1\columnwidth]{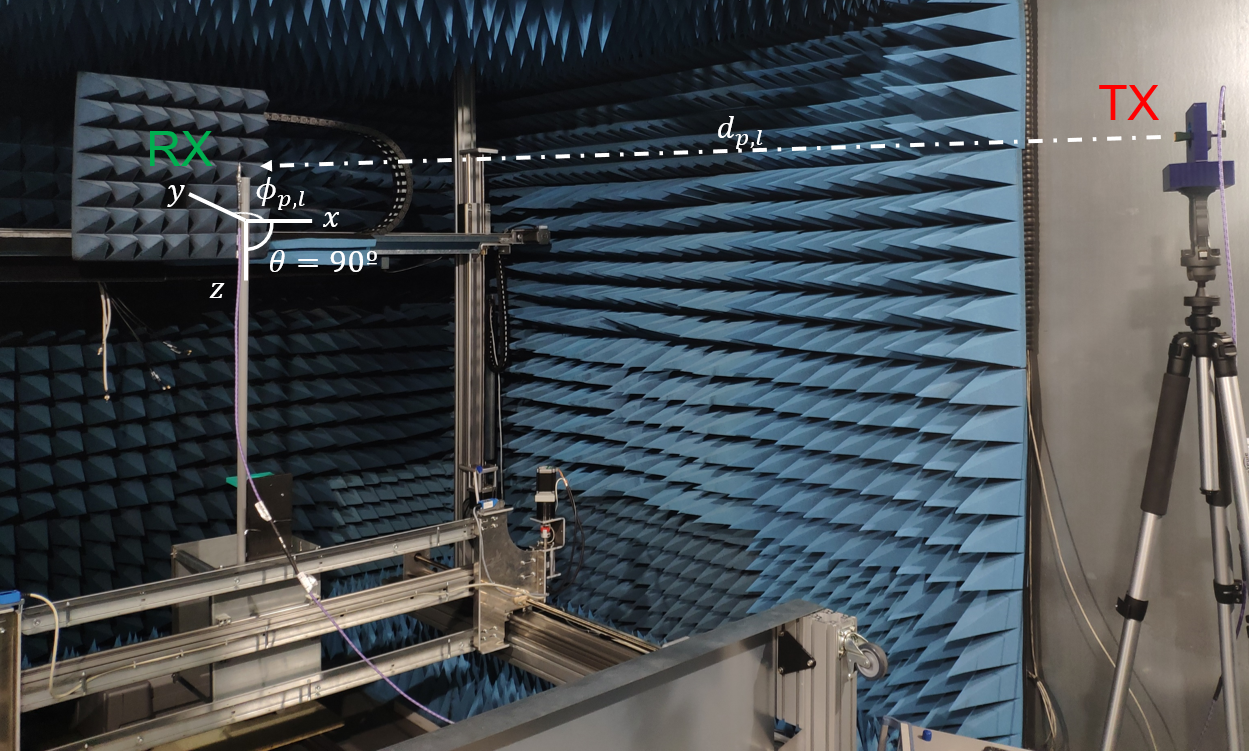}
	\caption{Photograph of the measurement setup. TX is fixed and RX forms the elliptical array due to the movement of the measurement system located at the bottom.} 
	\label{setup_camara}
\end{figure}

In order to validate the simulations and theoretical framework, multiple measurements have been carried out at the facilities of the University of Granada. These facilities consist of a semi-anechoic and semi-reverberation chamber whose dimensions are $5 \times 3.5 \times 3.5$ meters $(61.25$ m$^3)$. The measurements are performed in the semi-anechoic part, where multiple absorbers are found in the walls in order to avoid any reflection. Therefore, the Line-of-Sight (LoS) can be analyzed in the propagation channel between a transmitter (TX) and a receiver (RX). To recreate the simulation setup, TX is placed at a distance $d_{l}$ and azimuth angle $\phi_{l}$ with respect to the center of a certain ellipse, which is formed by a virtual array at RX. Fig. \ref{setup_camara} shows the measurement setup in the semi-anechoic chamber. The transmitter antenna is fixed, while the receiver antenna, located in the measurement system, can move in the XY plane positioned at the same height ($z = 179 \text{ cm}$) as the transmitter antenna. The measuring system allows a maximum displacement of 1 m in both x and y axes. The communication channel is acquired through a Vector Network Analyzer (VNA Rohde \& Schwarz ZVA67), which measures the scattering parameters up to 67~GHz. In order to prevent the effect of the coaxial cables in the communication channel, a Through — Open — Short — Match (TOSM) calibration is performed. Thus, $H_{p,l}$ includes the contribution of the propagation channel and the radiation pattern of the antennas in TX and RX. Particularly, TX is a standardized gain horn fed with a WR-15 waveguide-to-coaxial transition (Flann Kband antenna Model: \#25240-20). RX is a monopole antenna based on a 1.85~mm coaxial transition to free space, centered at 60~GHz, with a matching bandwidth higher than 8~GHz below $-10 \text{ dB}$, and omnidirectional radiation pattern for $\theta_l = 90\degree$ in the XY plane. The frequency range is chosen to be from 58 GHz to 62 GHz, for measurements and simulations, i.e., $B = 4$ GHz in the mmWave range. $K = 200$ frequency samples are acquired, providing 20~MHz frequency step. Given the bandwith $B$, the temporal and distance resolution are 0.25~ns and 7.5~cm respectively. Consequently, the maximum observable time and distance are 49.75~ns and 14.925~m.

\begin{figure}[t]
	\centering
	\subfigure[]{\includegraphics[width= 1\columnwidth]{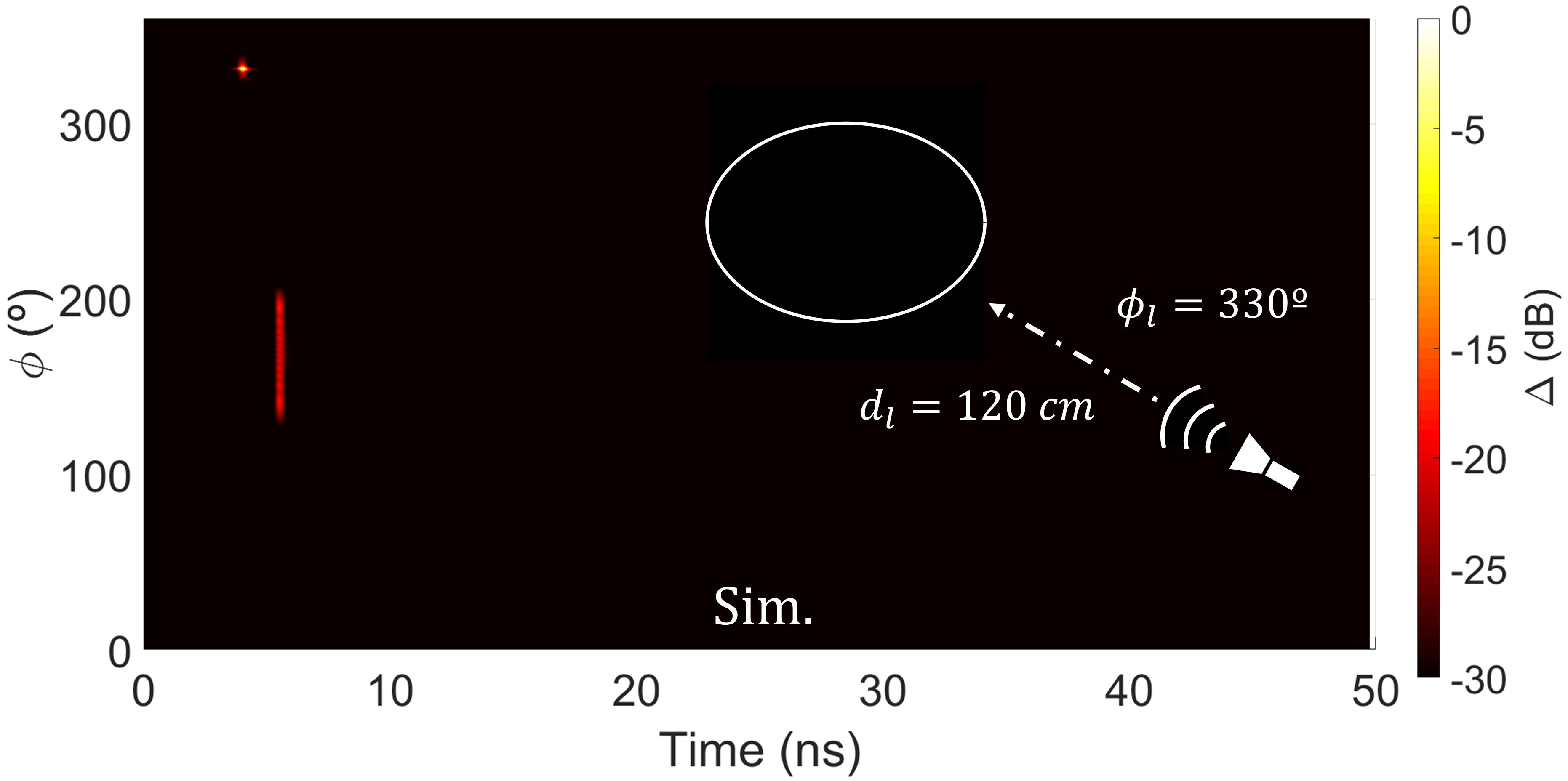}
	}
	\subfigure[]{\includegraphics[width= 1\columnwidth]{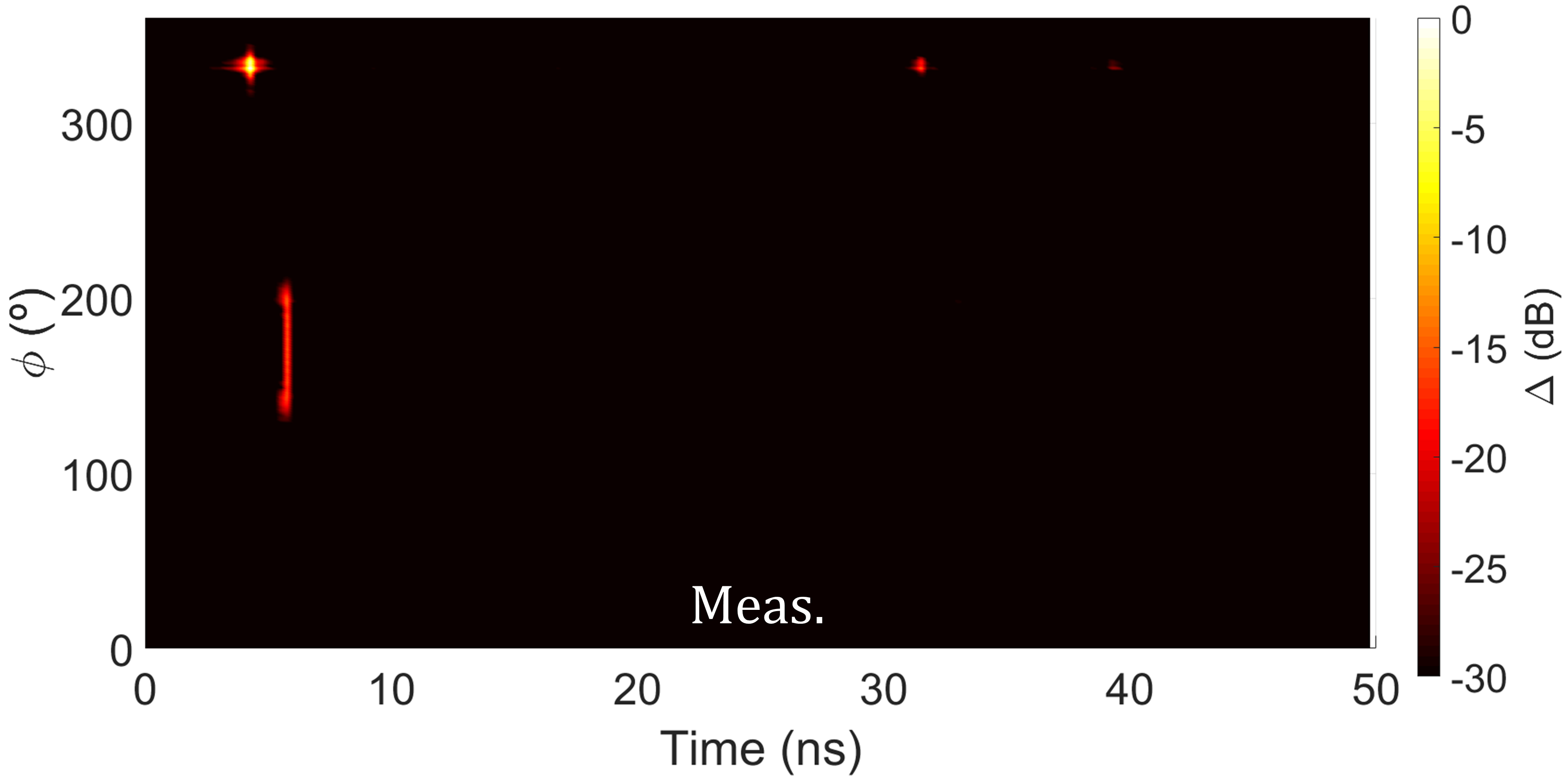}
	}
	\caption{DoA and ToA estimation for an incident wave at $\phi_{l} = 330\degree$ and $\tau_{l} = 4 \text{ ns } (d_{l} = 120 \text{ cm})$. (a) Simulation and (b) measurement in the semi-anechoic chamber.} 
	\label{elipse_10}
\end{figure}

In the first experiment, TX is placed at distance $d_{l} = 120$~cm and angle $\phi_{l} = 330\degree$. RX forms an ellipse with $a = 24.2$ cm, $\xi = 0.7$ and $\alpha = 0\degree$. $P = 720$ sensors are considered, for a maximum separation of 2.1 mm between sensors ($0.437\lambda$ for $f = 62$ GHz). Finally, the number of phase modes $M$ is set to 250. Figs. \ref{elipse_10}(a) and \ref{elipse_10}(b) show the DoA and ToA estimation for the simulated and measured cases, respectively. As expected, the estimation is maximized for $\tau_{l} = 4$ ns $(d_{l} = 120 \text{ cm})$ and $\phi_{l} = 330\degree$, with an artifact located at $\phi_{l} \pm 180\degree$. In Fig. \ref{elipse_10}(b), the estimation of DoA and ToA is slightly more spread due to measurement imperfections with respect to simulation. However, the maximum is still found at $\tau_{l} = 4.25$ ns $(d_{l} = 127.5 \text{ cm})$ and $\phi_{l} = 330\degree$. The variation of 7.5 cm is mainly due to the calibration process; namely, horn length (7.6 cm) and monopole transition (2 cm) are not initially considered. The distance resolution is 7.5 cm, thus the estimated value of $d_{l}$ is within the expected error band in measurement. Finally, note that two reflections due to back propagation in the semi-reverberation chamber can be found at 31.5 ns and 39.25 ns in Fig. \ref{elipse_10}(b).

\begin{figure}[t]
	\centering
	\includegraphics[width= 1\columnwidth]{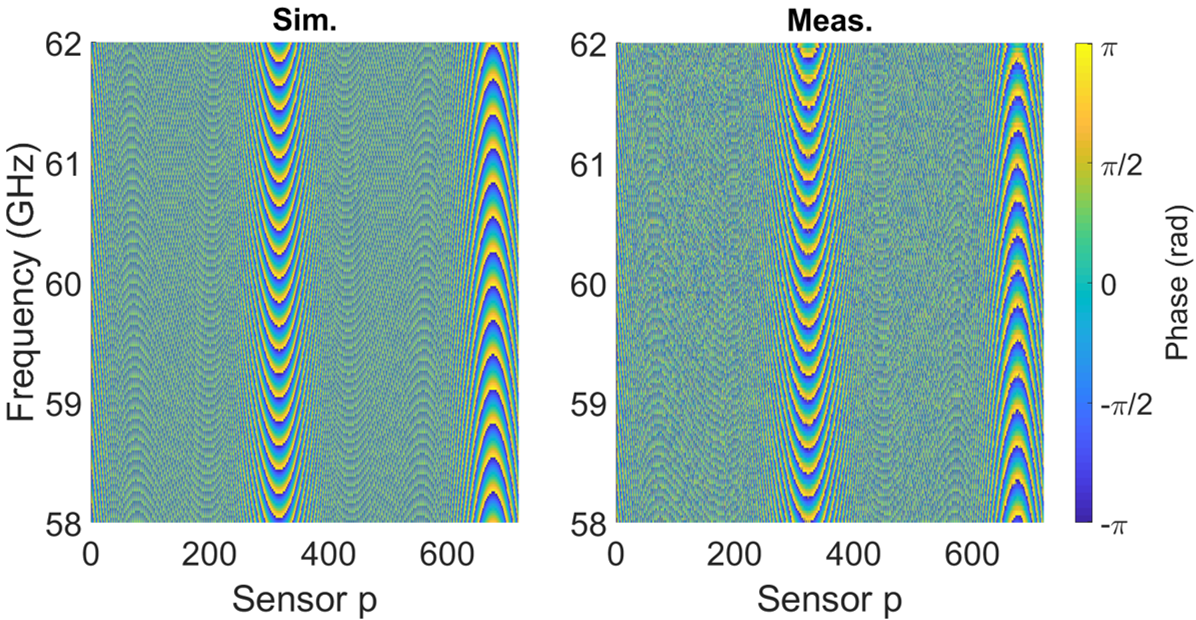}
	\caption{Simulated and measured $H_{p,l}(f)$ phase distribution when a single wave impinges at $\phi_{l} = 330\degree$ and $\tau_{l} = 4 \text{ ns}$. Ellipse parameters are $a = 24.2$ cm, $\xi = 0.7$, $\alpha = 0\degree$ and $P = 720$.} 
	\label{distribucion_fases_elipse_10}
\end{figure}

Fig. \ref{elipse_10} has shown a good agreement between the theoretical framework and the simulations proposed for this estimation method. As explained in Sec. II, this technique is based on a correct modeling of the incident signal on the $P$ sensors, i.e., $H_{p,l}$. Thus, phase-mode expansion $H_{m,l}$ and 2-D FFT lead to the joint DoA and ToA estimation. A proper modeling of the phase is fundamental in $H_{p,l}$. As an example, Fig.~\ref{distribucion_fases_elipse_10} illustrates the $H_{p,l}$ phase distribution for every sensor and frequency in the setup previously shown in Fig. \ref{elipse_10}. High similarity of the phase distribution in simulations and measurements can be observed. This fact demonstrates the proper modeling of the incident wave for elliptical cases in real measurements, thus yielding good joint DoA and ToA estimations. 

\begin{figure}[t]
	\centering
	\includegraphics[width= 1\columnwidth]{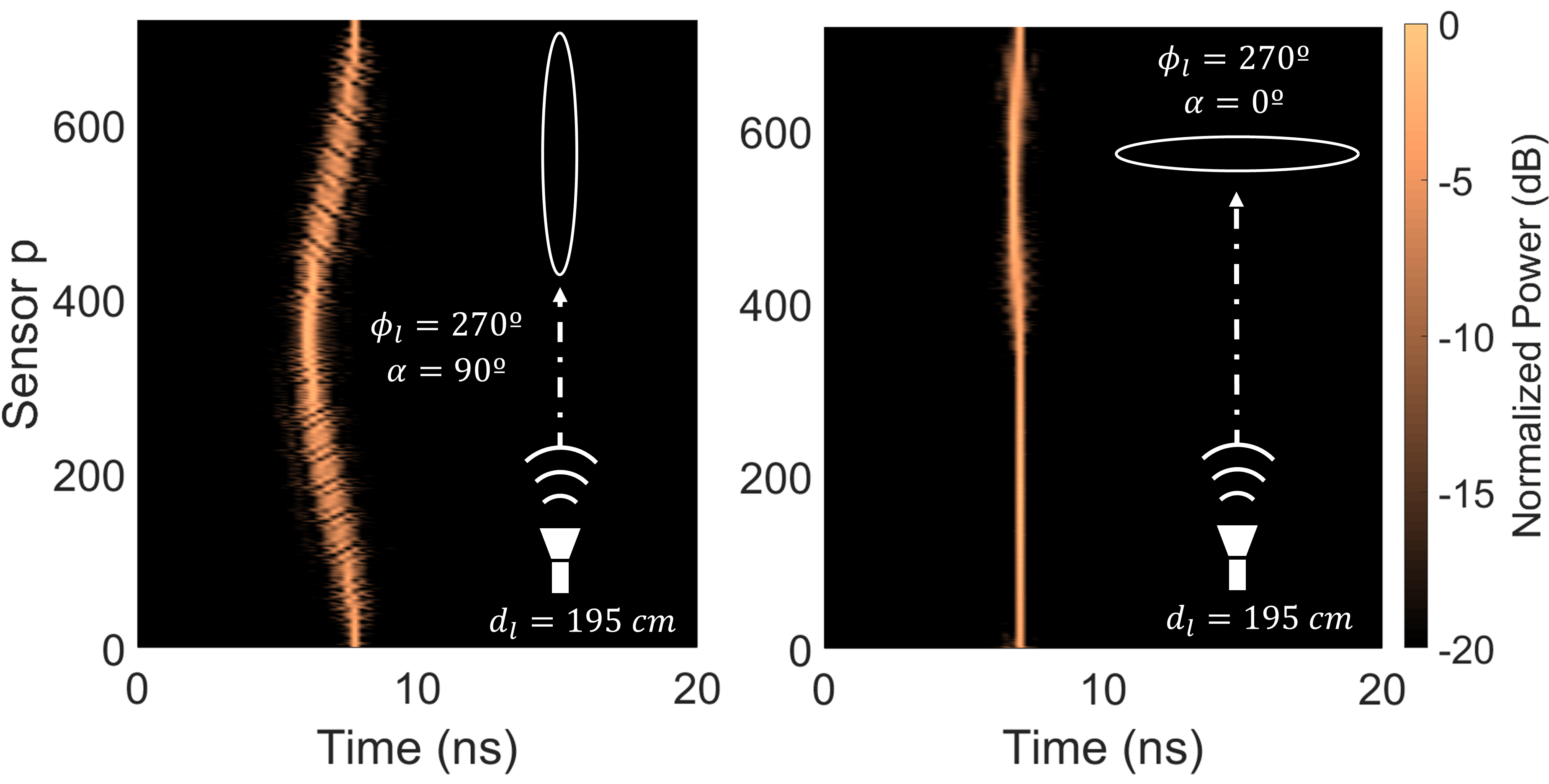}
	\caption{Normalized measured PDP for an incident wave at $\phi_{l} = 270\degree$ and $\tau_{l} = 6.5 \text{ ns}$ in two different setups: (left panel) rotated ellipse and (right panel) non-rotated ellipse.} 
	\label{CIR_concatenada}
\end{figure}

One of the features observed in the simulation is the high angular selectivity offered by highly flattened elliptical arrays. One way to explain this behavior is based on understanding how the wave reaches the array. For this purpose, two different setups have been measured. The first one has the TX positioned at a distance of $d_{l} = 195$ cm and an angle $\phi_{l} = 270\degree$. The RX is a virtual array with parameters:  $a = 24.2$ cm, $\xi = 0.99$, $\alpha = 90\degree$ and $P = 720$. The second setup is similar, except for the rotation angle $\alpha =0\degree $. Fig. \ref{CIR_concatenada} shows the normalized Power Delay Profile (PDP) for every sensor~$p$ in RX. This PDP is calculated as the square of the IFFT of the channel frequency response $H_{p,l}$ for each sensor. On the left panel, it can be seen that the wave impinges in the direction of the semi-major axis. Therefore, this wave travels sensor by sensor through the entire array. This fact generates the curvature observed in the PDP for the different sensor positions. On the right panel, the wave impinges in the direction of the semi-minor axis. In this case, for high eccentricities, the wave simultaneously reaches all the sensors of the array. This results in no curvature in the PDP, i.e, the measured channel for the different sensors is similar. In the first case, the curvature is unambiguous, since only this DoA generates such curvature. In the second case, the practical absence of curvature causes ambiguity in the estimation, since $\phi_{l} = 90\degree$ would generate exactly the same PDP. Hence, it can be concluded by the measurement study that this technique takes advantage of the curvature of the elliptical array, thus justifying the curves simulated in Fig. \ref{delta_elipses}. As it was noted in Sec. III.A, the selectivity is an attribute of the sensor arrangement.

\begin{figure}[t]
	\centering
	\includegraphics[width= 1\columnwidth]{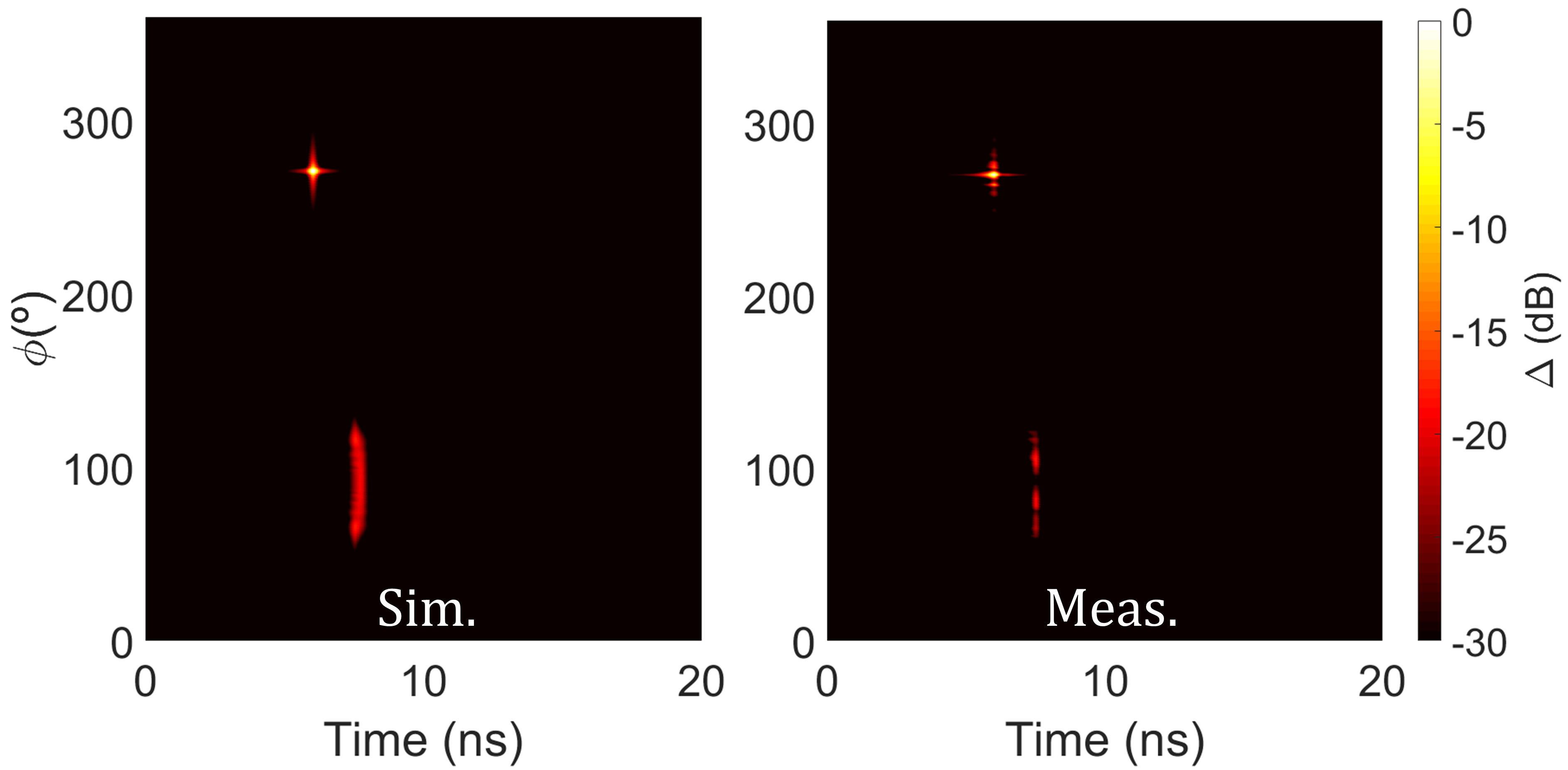}
	\caption{Simulated and measured DoA and ToA  estimation for an incident wave at $\phi_{l} = 270\degree$ and $\tau_{l} = 5.66 \text{ ns}$ when the elevation angle is $\theta_l =\nolinebreak70\degree$.} 
	\label{medidas_elevacion_70}
\end{figure}

Sec. III.C presented the technique performance for elevation angles $\theta_l \neq 90 \degree$. In order to validate previous simulations, a setup with $\theta_l =\nolinebreak70\degree$ is measured. TX antenna is placed at height $z = 120 \text{ cm}$ with $\phi_{l} = 270\degree$, and the elliptical array is kept at $z = 179 \text{ cm}$. The distance in the XY plane between TX and the center of the elliptical array is 160 cm. Therefore, by applying basic trigonometry, $d_{l} =\nolinebreak170 \text{ cm } (\tau_{l} = 5.66 \text{~ns})$ and $\theta_l$ turns out to be $70\degree$. RX parameters are $a = 24.2$ cm, $\xi = 0.7$, $\alpha = 90\degree$, $P = 720$ and $M = 250$. Fig.~\ref{medidas_elevacion_70} shows the simulated and measured joint estimation of the DoA and ToA for the previous scenario. The measured scenario matches the simulation with $270\degree$ DoA and 6 ns ToA estimation. The artifacts appear 16 dB below the estimation, showing a good agreement with the simulation prediction (see $\xi = 0.7$ in Fig.~\ref{fig_elevacion}).

\begin{figure}[b]
	\centering
	\includegraphics[width= 1\columnwidth]{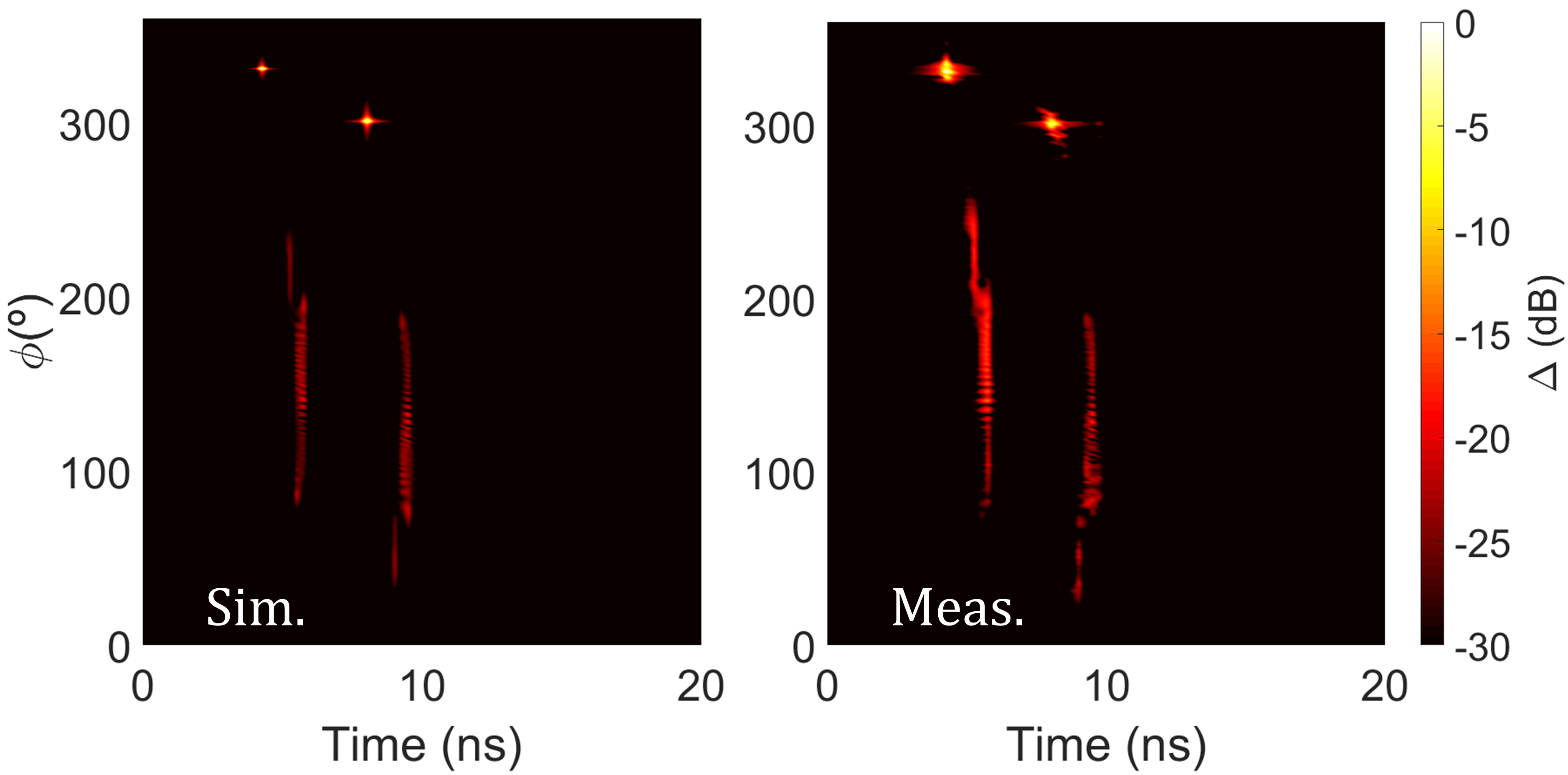}
	\caption{Simulated and measured DoA and ToA  estimation for a multipath component scenario and concentric elliptical arrays. The first ray reaches RX for $330\degree$ DoA and $4 \text{ ns}$ ToA, and the second ray for $300\degree$ DoA and $8 \text{ ns}$. Two ellipses with $\xi = 0.7$ and different rotation angle are considered.} 
	\label{dos_rayos}
\end{figure}

For the sake of completeness of the experimental study, it will be shown that the use of this technique is valid for i) multipath environments, and ii) concentric elliptical arrays. For that purpose, i) two measured rays are combined in the semi-anechoic chamber as the sum of the frequency response of each of the rays incident at the $p$-th sensor [see eq. \eqref{Ym}]. The first wave reaches the center of the array from $330\degree$ and 4~ns (120 cm) delay, while the second wave impinges the ellipse from $300\degree$ azimuth angle and 8~ns (240 cm) delay. ii) Two ellipses are combined to form a concentric elliptical array. The phase-mode response from the sum of the two previous waves is summed to create the new arrangement [see eq. \eqref{Hml_psi2}]. The parameters for the first elliptical array are $a = 24.2$~cm, $\xi =\nolinebreak0.7$, $\alpha = 0\degree$, $P = 720$ and $M = 250$, while the second array is characterized by the same parameters except for $\alpha = 90\degree$. Fig.~\ref{dos_rayos} shows the joint estimation for the multipath scenario and the superposition of ellipses. Both waves can be clearly depicted in the simulations and measurements for the correct values of DoA and ToA. A good estimation is seen, which validates the proposed approach and the experimental setup for usage in multipath environments in wireless communication links.

In general, it should be stated that the level of the main artifacts varies as multiple concentric arrays of different geometries are considered. Flattened elliptical arrays present angular selectivity, normally in angular regions near  the semi-major axis (see Fig. 4). This fact provokes that the level of artifacts either decreases or increases depending on the location of the DoA. In the case that the semi-major axis of the elliptical array is aligned with the DoAs, the level of the artifacts decreases (the metric $\Delta$ increases). On the other hand, if the DoAs are not aligned with the semi-major axis, the level of the artifacts is expected to increase (the metric $\Delta$ decreases). More than a limitation of the proposed method, the modification of $\Delta$ as a consequence of the orientation of the array should be considered an inherent property of flattened elliptical and linear arrays.

\section{\label{sec:Conclusions} Conclusions}

This work proposes a technique for joint DoA and ToA estimation based on ultra-wideband elliptical arrays. The theoretical framework introduces a generalization of frequency-independent beamformers based on uniform circular arrays. This new approach is a generalization not only valid for elliptical arrays, but also feasible for circular and linear arrays. The geometry of the elliptical array provides new degrees of freedom compared to the circle, where only the radius can be modified. The ellipses, besides tuning the semi-major axis, also allow to adjust their eccentricity or rotation angle. These degrees of freedom, together with the superposition of concentric ellipses, result in pseudo-random sensor geometries. This novel concept avoids the problem of being limited to a specific set of geometries, being able to adapt the array to several arrangements in a real deployment.

The analysis of the technique through simulations has shown good joint DoA and ToA estimations, with artifacts appearing orders of magnitude below the main component. The study has been performed for the whole range of azimuth angles $\phi_l$, as well as for multiple time-of-arrival $\tau_l$, and three frequency bands within the mmWave range: 28-30~GHz, 39.5-43.5~GHz and 58-62~GHz. Additionally, the effect of modifying the rotation angle $\alpha$, eccentricity $\xi$ and elevation angles $\theta_l$ on the joint estimation has been analyzed. Through these degrees of freedom, arrangements of nine geometries with pseudo-random array distribution have been simulated. One step further, an independent and uncorrelated bivariate normal distribution is included in the sensor position in order to recreate a random grid. The results show that even including this random distribution, the estimation can be properly performed.

Finally, the simulations have been validated through multiple measurements in a semi-anechoic chamber at 58-62 GHz frequency band. DoA and ToA estimation has been initially carried out for single ellipses and single-path environments. Then, this has been extended to multipath scenarios and concentric elliptical arrays, simultaneously. Both the measured frequency response, phase distributions and the joint estimation of the DoA and ToA agree with those predicted by the simulations. These experimental results validate the theoretical framework, thus providing a method which gives accurate estimations for a large number of sensor arrangements and arbitrary geometries.

\ifCLASSOPTIONcaptionsoff
  \newpage
\fi

\end{document}